\documentclass[11pt]{article}
\usepackage[margin=1in]{geometry}
\usepackage[T1]{fontenc}
\usepackage{lmodern}
\usepackage[numbers,sort&compress]{natbib}
\usepackage{graphicx}
\usepackage{amsmath,amssymb,amsthm}
\usepackage{algorithm}
\usepackage{algpseudocode}
\usepackage{booktabs}
\usepackage{tabularx}
\usepackage{threeparttable}
\usepackage{array}
\usepackage{multirow}
\usepackage{makecell}
\usepackage{xcolor}
\usepackage{url}
\usepackage{hyperref}

\graphicspath{{figures/}{../figures/}}
\theoremstyle{remark}

\newcommand{\papertablesize}{\scriptsize}
\newcommand{\prioritypolicy}{Q75 Margin-Gated Floor policy}

\title{Decision Support for Marketplace Policies under Incomplete Evidence: From Replay to Launch Readiness}
\author{Prashant Shekhar and Caroline Howard}
\date{\small\textit{Department of Mathematics}\\\textit{Embry-Riddle Aeronautical University, Daytona Beach, FL, USA}}

\begin{document}
\maketitle
\begin{abstract}
Marketplace platforms routinely evaluate pricing and allocation policies using logged observational data, yet strong offline performance does not imply that a policy is safe to deploy. In real-time bidding (RTB) marketplaces, reserve-price and floor-policy changes affect not only revenue but also fill, advertiser value, budget pacing, and competition across auctions, creating complex feedback and interference. The central problem is therefore not to estimate whether a policy improves an offline metric, but to determine whether the available evidence justifies direct launch or only further validation. We propose a support-aware decision-support system (DSS) that explicitly quantifies favorable and unfavorable offline evidence for candidate policies, while making a distinction between \textit{promising} and \textit{actionable} insights. The framework integrates a formal replay estimand, support-aware off-policy evaluation (OPE), conservative lower-bound ranking, multi-sided guardrails, out-of-time validation, sensitivity analysis, and interference-aware validation design into a claim-preserving decision pipeline that outputs a launch-readiness classification rather than a single performance estimate. Applying the framework to iPinYou-style RTB logs, we identify a margin-gated floor policy as the leading candidate, with a 47.7\% replay yield lift, a 45.8\% conservative lower-tail lift, and stable out-of-time performance (43.9\% lift with full retention of impressions and value proxies). However, the framework does not recommend direct launch. A decision-rule ablation shows that replay-only, OPE-only, guardrail-only, and holdout-only pipelines all select the same policy but would incorrectly recommend deployment, leaving critical causal assumptions unresolved. In contrast, the proposed DSS framework selects the same policy but changes the action to online validation, reflecting missing evidence on propensities, bidder response, and interference. Overall, the contribution is a reproducible DSS protocol that prevents decision overclaim under partial identification, and converts offline policy evaluation into an auditable, action-oriented recommendation.

\end{abstract}


\section{Introduction}

Data science systems increasingly support high-stakes platform decisions, such as which ranking policy to deploy, which pricing rule to test, which marketplace intervention to hold, and which results should be escalated to production. In computational advertising, these decisions are particularly challenging because learning systems operate within dynamic auction environments that involve users, advertisers, and feedback loops \citep{bottou2013counterfactual}. In real-time bidding (RTB) marketplaces, reserve-price and floor-policy changes affect multiple outcomes simultaneously, including platform yield, fill rate, advertiser value, budget pacing, and competition across auctions. As a result, even seemingly simple pricing interventions can propagate through the marketplace in complex and difficult-to-predict ways.

The central problem is therefore not merely to estimate whether a candidate policy improves an offline metric, but to determine whether the available evidence is sufficient to support a production decision. In practice, platforms must distinguish among four operational actions: (i) direct launch, (ii) online validation, (iii) hold, or (iv) redesign. This distinction is critical because offline estimates, whether obtained through replay or statistical evaluation, typically rely on assumptions that do not fully capture bidder behavior, market equilibrium, or interference across auctions. A decision-support system (DSS) for marketplace policies must therefore explicitly map incomplete and heterogeneous evidence into an appropriate operational action.

This issue is not only conceptual but empirically consequential. In our case study, multiple simplified decision rules including replay-only, replay with guardrails, off-policy evaluation (OPE) mean-only, OPE lower-tail-only, and holdout replay identify the same reserve/floor policy as optimal. However, when used as standalone decision systems, each would recommend direct launch based on incomplete logged evidence. In contrast, the full DSS framework proposed in this paper selects the same policy but changes the action to online validation, because key sources of uncertainty such as production propensities, bidder response, and interference remain unresolved. This result highlights the central contribution of the paper: the goal is not to change which policy appears best, but to change what the platform is justified in doing with that evidence.

Our approach builds on and integrates several strands of prior research. Reserve-price optimization and RTB studies establish that floor policies can materially affect platform outcomes but depend on bidder behavior and market structure \citep{zhang2014ipinyou,cai2017rtb}. The off-policy evaluation literature provides estimators, such as inverse propensity weighting and doubly robust methods, for evaluating policies from logged data under support assumptions \citep{bottou2013counterfactual}. Marketplace experimentation research shows that interference arising from shared budgets and pacing can invalidate naive randomized experiments and motivates designs such as switchbacks \citep{bottou2013counterfactual}. Finally, the DSS literature emphasizes interpretable, decision-relevant analytics that connect model outputs to operational consequences \citep{coussement2021interpretable,abedin2024explainable,decorte2023costsensitive}. Despite these advances, a critical gap remains. In particular, existing components provide partial answers, but do not specify how to combine them into a coherent decision protocol under incomplete identification.

To address this gap, we propose a support-aware decision-support framework for reserve/floor-policy evaluation in RTB marketplaces. The framework integrates auction replay, support-aware OPE diagnostics, conservative lower-bound ranking, multi-sided guardrails, sensitivity analysis, and interference-aware validation design into a unified decision pipeline. Rather than producing a single policy estimate, the framework outputs a launch-readiness classification together with the evidence and assumptions that justify it. Conceptually, the framework is a \emph{claim-preserving decision map} stating that each diagnostic contributes only to the claims it can support, and the final recommendation is constrained by the weakest unresolved assumption.

The contributions of the paper can be summarized as follows:

\begin{itemize}
  \item \textbf{A claim-preserving decision framework for marketplace policy evaluation}: We develop a formal decision map and algorithm that convert heterogeneous offline evidence into operational actions (launch, validate, hold, or redesign) while preserving the causal meaning of replay, OPE, guardrails, response sensitivity, and interference diagnostics.

  \item \textbf{A formal interpretation of auction replay as a bounded mechanical estimand}: We clarify the claim boundary of static replay and show how replay-based gains must be interpreted under fixed bidder behavior, complemented by explicit bidder-response sensitivity analysis.

  \item \textbf{Support-aware OPE diagnostics for decision-making}: We extend standard OPE evaluation by incorporating support diagnostics, effective sample size, clipping sensitivity, and conservative lower-tail ranking, and show how these translate into validation-readiness rather than direct-launch claims.

  \item \textbf{An integrated offline-to-online validation protocol}: We connect offline policy evaluation to experimental design by introducing an explicit sequence from replay and OPE diagnostics to shadow logging and interference-aware switchback experiments.

  \item \textbf{An empirical demonstration of decision overclaim and its correction}: Using iPinYou-style RTB logs \citep{zhang2014ipinyou,cai2017rtb}, we show that multiple simplified evaluation pipelines select the same policy but incorrectly recommend direct launch, while the proposed DSS framework selects the same policy and correctly recommends online validation. The selected policy achieves a 47.7\% replay lift, a 45.8\% conservative lower-tail lift, and a 43.9\% out-of-time replay lift, yet remains not launch-ready due to unresolved causal uncertainties, quantified through a decision chaining strategy.

\end{itemize}

%In policy names, Q25, Q50, and Q75 denote the 25th-, 50th-, and 75th-percentile floor thresholds, respectively.

\section{Related Work}

This paper draws on and integrates four strands of literature: 

\begin{enumerate}
    \item \textbf{Reserve prices and RTB marketplaces}: A large body of work studies how reserve prices and floor policies affect auction outcomes and platform revenue. Early empirical and theoretical studies show that reserve prices can substantially influence allocation and payments in advertising auctions \citep{ostrovsky2011reserve,yuan2014reserve}, while more recent work considers learning-based and data-driven approaches to reserve optimization \citep{kalra2023reserve,choi2025reserve}. Complementary RTB research examines related components of marketplace dynamics, including constrained display-ad allocation, bid-landscape forecasting, sequential bidding, and budget pacing \citep{chen2011rtb,cui2011bidlandscape,cai2017rtb,balseiro2019learning}. Collectively, these studies establish that floor policies can materially affect platform outcomes, but also depend critically on bidder behavior, auction format, and demand constraints. However, they typically focus on policy optimization or estimation, rather than on the decision of whether a policy is ready for deployment.

    \item \textbf{Off-policy evaluation and counterfactual learning}:
    A second line of work provides tools for evaluating candidate policies using logged data. Replay-based evaluation for contextual bandits demonstrates how randomized logs can be used to estimate policy performance without exposing users to untested decisions \citep{li2011unbiased}. More generally, OPE methods such as inverse propensity weighting (IPW) and doubly robust (DR) estimation combine propensity modeling and outcome regression to improve statistical efficiency and robustness \citep{dudik2011doubly,jiang2016doubly}. Counterfactual risk minimization extends these ideas to policy learning from logged feedback \citep{swaminathan2015counterfactual}, while subsequent work on adaptive weighting and shrinkage highlights the importance of stability under limited support \citep{wang2017adaptive,su2020shrinkage,zhan2021adaptive}. Benchmark datasets and pipelines further emphasize reproducibility and diagnostic reporting \citep{saito2021openbandit}. Although this literature provides powerful estimators, it largely treats evaluation as a statistical problem and does not specify how estimator outputs should be translated into operational decisions under support limitations.

    \item \textbf{Marketplace experimentation under interference}:
    A third strand of research addresses the challenges of causal inference in two-sided marketplaces. When advertisers share budgets, pacing constraints, or inventory, standard randomized experiments can be biased due to interference across units \citep{holtz2020interference,li2022interference,johari2022experimental,bajari2023marketplace}. To address these issues, the literature proposes alternative experimental designs, including cluster-based randomization and switchback experiments, which align treatment assignment with marketplace structure and temporal dynamics \citep{candogan2023cluster,bojinov2022switchback,ni2025switchback,chen2025surrogate}. This work demonstrates that offline evaluation is insufficient for identifying live marketplace effects, but typically treats experiment design as a separate stage following policy selection.

    \item \textbf{Decision support systems}:
    Finally, the DSS literature provides the overarching design objective. Decision-support systems aim to produce interpretable, decision-relevant outputs that connect data analysis to actionable choices. Recent work emphasizes transparency, explainability, and alignment with operational objectives \citep{coussement2021interpretable,abedin2024explainable}, as well as cost-sensitive and decision-aware modeling \citep{decorte2023costsensitive}. However, there is limited guidance on how to construct auditable decision artifacts for causal marketplace interventions when identification is incomplete.
\end{enumerate}

\begin{table}[!t]
\centering
\begin{threeparttable}
\caption{Positioning relative to prior research.}
\label{tab:literaturegap}
\papertablesize
\renewcommand{\arraystretch}{1.18}
\begin{tabularx}{\linewidth}{
  >{\raggedright\arraybackslash}p{0.30\linewidth}
  >{\raggedright\arraybackslash}p{0.24\linewidth}
  >{\raggedright\arraybackslash}p{0.22\linewidth}
  >{\raggedright\arraybackslash}X}
\toprule
Research stream and representative references & What it contributes & Remaining launch gap & Addition in this paper \\
\midrule
Reserve prices and RTB auctions. Representative references: \citep{ostrovsky2011reserve,yuan2014reserve,kalra2023reserve,choi2025reserve}; RTB foundations \citep{chen2011rtb,cui2011bidlandscape,cai2017rtb,balseiro2019learning}. & Mechanisms and empirical evidence for floor optimization & Replay gains can be mistaken for live causal launch effects & Formal replay claim boundary plus bidder-response sensitivity \\
Logged bandit and OPE methods. Representative references: \citep{li2011unbiased,dudik2011doubly,swaminathan2015counterfactual,wang2017adaptive,su2020shrinkage,zhan2021adaptive,saito2021openbandit}. & Estimators for policy evaluation from logged data & Estimator output may be used without support, clipping, or guardrail gates & Support-aware OPE scorecard and conservative lower-bound ranking \\
Marketplace experimentation. Representative references: \citep{holtz2020interference,li2022interference,johari2022experimental,bajari2023marketplace,candogan2023cluster,bojinov2022switchback,ni2025switchback,chen2025surrogate}. & Designs that address interference and platform state & Offline policy evaluation and online validation are often treated separately & Offline-to-online validation sequence with switchback readiness checks \\
DSS and interpretable decision support. Representative references: \citep{coussement2021interpretable,decorte2023costsensitive,abedin2024explainable}. & Emphasis on transparent, decision-relevant analytics & Less guidance on causal marketplace launch decisions under incomplete identification & Auditable recommendation artifact: launch, validate, hold, or redesign \\
\bottomrule
\end{tabularx}
\end{threeparttable}
\end{table}

Table~\ref{tab:literaturegap} summarizes the relationship between these literatures. For example, replay methods quantify mechanical policy effects, OPE methods estimate counterfactual performance under support assumptions, and experimentation designs address interference in live systems. The key gap is that these components are typically applied in isolation. As a result, platforms lack a principled method for combining partial evidence into a single operational decision.

The present paper addresses this gap by introducing a claim-preserving decision framework that integrates these components into a unified pipeline. In this framework, replay is interpreted as a bounded mechanical estimand, OPE serves as a support-aware diagnostic, and interference theory determines the requirements for online validation. The resulting output is not a point estimate but a launch-readiness classification supported by explicit evidence and assumptions.

This distinction is central to the paper’s novelty. A replay-based analysis can answer whether a policy would have improved revenue under fixed bidder behavior. An OPE analysis can assess estimator performance under a target policy and logging distribution. An experimentation study can determine how to randomize safely in a marketplace. In contrast, this paper addresses the operational question that connects these components. In formal terms, given incomplete and heterogeneous evidence, what action should the platform take next? The empirical contribution is therefore evaluated not by improved estimation alone, but by whether the integrated framework changes the resulting decision relative to replay-only or estimator-only workflows.

\section{Decision Setting}

We study decision-making in a repeated display-ad auction marketplace. For each impression opportunity, a platform or publisher exposes inventory through an exchange, advertisers submit bids, and the auction mechanism determines allocation and payment. A reserve price or floor policy specifies the minimum acceptable payment for the impression. Changing this floor alters both allocation and pricing. In other words, a higher floor can increase payments on retained impressions, but can also reduce the probability of sale by excluding lower bids. The platform therefore faces a fundamental \emph{revenue--liquidity tradeoff}: more aggressive floors may increase yield per retained impression while reducing fill, advertiser reach, and downstream engagement.

Formally, let $i=1,\ldots,n$ index logged auction opportunities. Each observation consists of a context $X_i$, a logged floor $f_i^0$, an observed bid or winning-bid proxy $b_i$, a realized payment $p_i$ when the auction clears, and a filled-impression indicator $D_i^0 \in \{0,1\}$. The context $X_i$ may include time, exchange, region, device, slot, advertiser, campaign, and user or page attributes. The baseline outcome of interest is the realized yield contribution $Y_i^0 = D_i^0 p_i$. A candidate floor policy $\pi$ maps context to a proposed floor, $f_i^\pi = \pi(X_i)$. In this paper, the policy class is deliberately restricted to deterministic and auditable rules such as quantile floors, margin-based adjustments, and hybrid thresholds to ensure that policy behavior is interpretable and operationally verifiable.

In this section we describe the structural features of the decision setting of interest, and explain how it differs from standard supervised learning or binary treatment problems.

\paragraph{Partial feedback and identification limits:}
The first complication is partial feedback. When an auction is not filled ($D_i^0 = 0$), the platform does not observe the payment that would have occurred under a lower floor or under different bidder behavior. This missing counterfactual limits what can be identified from logged data. To maintain a transparent and well-defined estimand, the replay analysis is restricted to non-decreasing floor policies ($f_i^\pi \geq f_i^0$). Under this restriction, candidate policies can only retain or drop observed filled impressions; they cannot create new filled outcomes that were not observed in the log. This yields a mechanically interpretable replay estimand, but also highlights a key limitation that replay captures only fixed-behavior effects and cannot recover outcomes that depend on unobserved counterfactual participation.

\paragraph{Multi-action decision problem:}
The second complication is that the platform’s objective is not to estimate a treatment effect, but to choose among multiple operational actions. The output of the analysis must support one of four decisions: {launch}, {validate online}, {hold}, or {redesign}. These actions differ not only in expected performance but also in the strength of evidence required. Table~\ref{tab:decisionstates} summarizes the corresponding decision states. Crucially, a policy can exhibit large positive offline lift yet still be unsuitable for direct launch if key sources of uncertainty such as bidder response or interference remain unresolved. This distinction elevates the problem from estimation to decision support.

\begin{table}[!t]
\centering
\begin{threeparttable}
\caption{Operational decision states for a reserve/floor-policy candidate.}
\label{tab:decisionstates}
\papertablesize
\renewcommand{\arraystretch}{1.12}
\begin{tabularx}{\linewidth}{
  >{\raggedright\arraybackslash}p{0.18\linewidth}
  >{\raggedright\arraybackslash}p{0.36\linewidth}
  >{\raggedright\arraybackslash}X}
\toprule
Action & Evidence threshold & Typical interpretation \\
\midrule
Launch & Strong point and lower-bound evidence, credible propensities, passed guardrails, and no unresolved interference or response risk & Evidence is sufficient for production rollout under pre-specified monitoring. \\
Validate online & Strong offline priority, but missing live propensities, bidder response, or interference-safe evidence & Policy deserves shadow logging, switchback testing, or limited ramp validation. \\
Hold & Mixed evidence, unstable support, or unresolved guardrail risk & Policy should not consume launch capacity until diagnostics improve. \\
Redesign & Weak support, dominated tradeoff, harmful segment effects, or unmeasured outcomes & Policy class, logging design, or measurement contract must be revised. \\
\bottomrule
\end{tabularx}
\end{threeparttable}
\end{table}

\paragraph{Multi-sided objectives and guardrails:}
Although yield per opportunity is the primary business metric, the decision problem is inherently multi-sided. A policy that increases revenue may simultaneously degrade marketplace quality or concentrate harm in specific segments. Accordingly, evaluation must incorporate additional quantities, including retained impression share, advertiser value proxies, click and conversion proxies, and segment-level outcomes. These metrics serve different roles such as yield and retention characterize the mechanical tradeoff; engagement and value proxies guard against quality degradation; and segment diagnostics ensure that aggregate gains do not mask localized harm. In contrast, support and propensity diagnostics do not measure welfare directly; they quantify the reliability of the evidence used to justify a decision.

\paragraph{Three distinct estimands:}
A central conceptual distinction in this setting is between three different estimands. The {static replay value} measures what would have occurred if candidate floors had been applied to the logged auctions under fixed bids and participation. The {off-policy value} measures what can be estimated under a logging policy with known propensities and sufficient overlap. The {live launch effect} captures the true marketplace impact, including bidder adaptation, budget reallocation, and dynamic interactions across auctions. These quantities are not interchangeable. The core methodological risk is to interpret replay or OPE estimates as if they identified the live effect.

\paragraph{Interference and marketplace dynamics:}
A central challenge in this decision setting is that auction outcomes are not independent across time. As illustrated in fig.~\ref{fig:interference_framework}(a), where the \emph{solid arrows} represent the offline evaluation path, where candidate policies are assessed using replay and OPE under a fixed marketplace state. This path supports policy selection from logged data, but it does not capture how the marketplace itself would respond to the policy. In contrast, the \emph{dashed arrows} represent the dynamics that arise after deployment. When a policy is implemented, its effects propagate through shared advertiser state, including budgets, pacing, and competition, which in turn influence future auctions. For example, raising floors in early auctions may exhaust advertiser budgets or change pacing behavior, reducing bidding intensity in later auctions. These feedback effects are intrinsic to the live marketplace but are not observed in static logs.

This creates a fundamental decision gap. The platform must choose a policy using evidence generated under the solid-path assumptions, while the true impact of that policy is governed by the dashed-path dynamics. As a result, offline estimates identify promising candidates but do not, by themselves, justify deployment. In this framework, strong offline results are therefore treated as \emph{validation-priority evidence}, and the decision problem becomes how to transition from offline selection to reliable online evaluation. Resolving this gap requires interference-aware experiments, such as the switchback design described in fig.~\ref{fig:decision_validation_sequence}(b).

\section{Methodology: Support-Aware Launch-Readiness Framework}

This section describes the operational methodology introduced by the paper. The goal is to turn logged auction data into a decision about whether a candidate reserve/floor policy should be launched, validated online, held, or redesigned. The methodology is intentionally more restrictive than a pure policy-optimization routine. It does not ask only which policy maximizes a point estimate, rather it asks which claims are identified by the available data, which claims depend on untested assumptions, and which next action follows from that evidence.

\begin{figure}[!t]
  \centering
  \begin{minipage}[t]{0.49\textwidth}
    \centering
    \includegraphics[width=\linewidth,height=0.38\textheight,keepaspectratio]{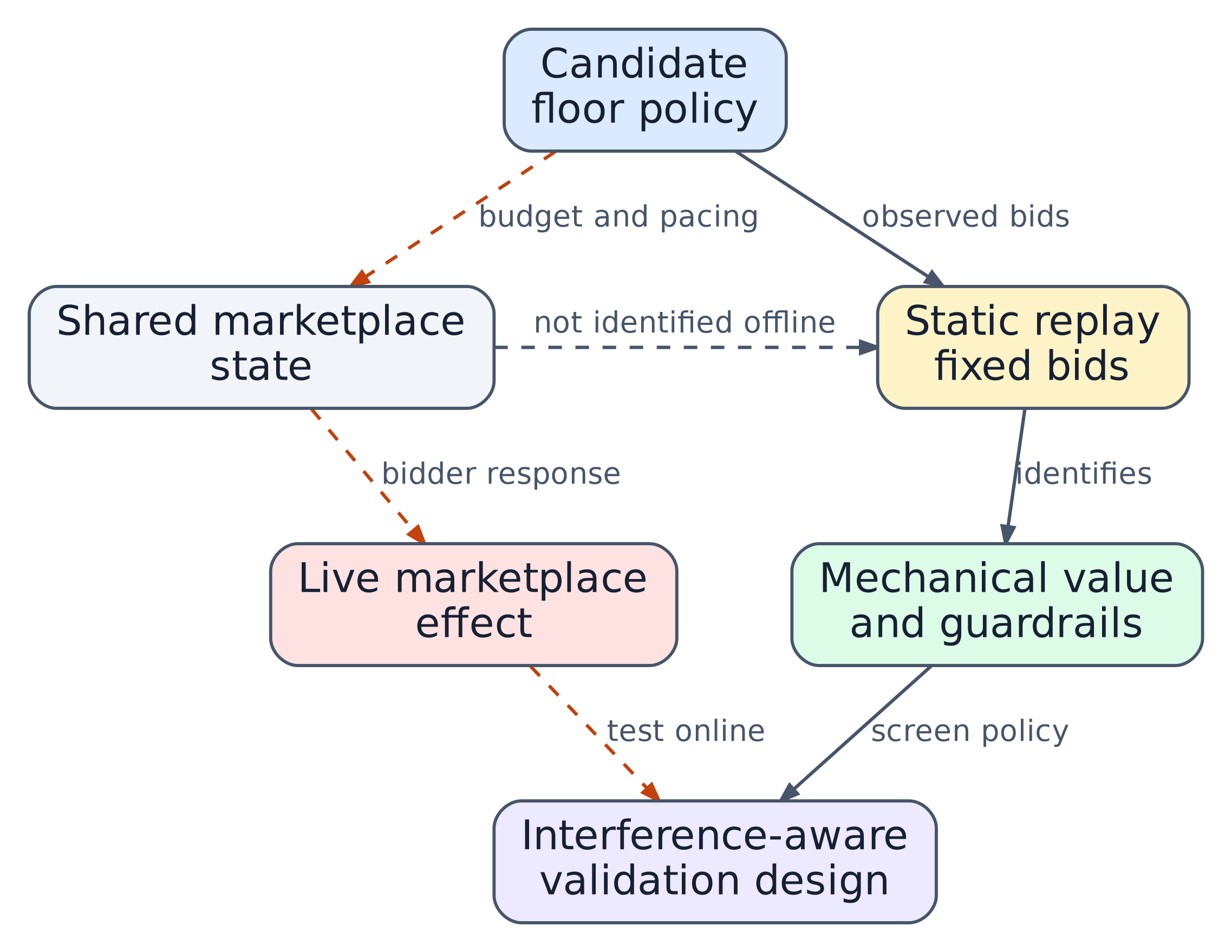}
    \vspace{0.4em}
    {\small\textbf{(a)} Marketplace interference paths.}
  \end{minipage}\hfill
  \begin{minipage}[t]{0.51\textwidth}
    \centering
    \includegraphics[width=\linewidth,height=0.26\textheight,keepaspectratio]{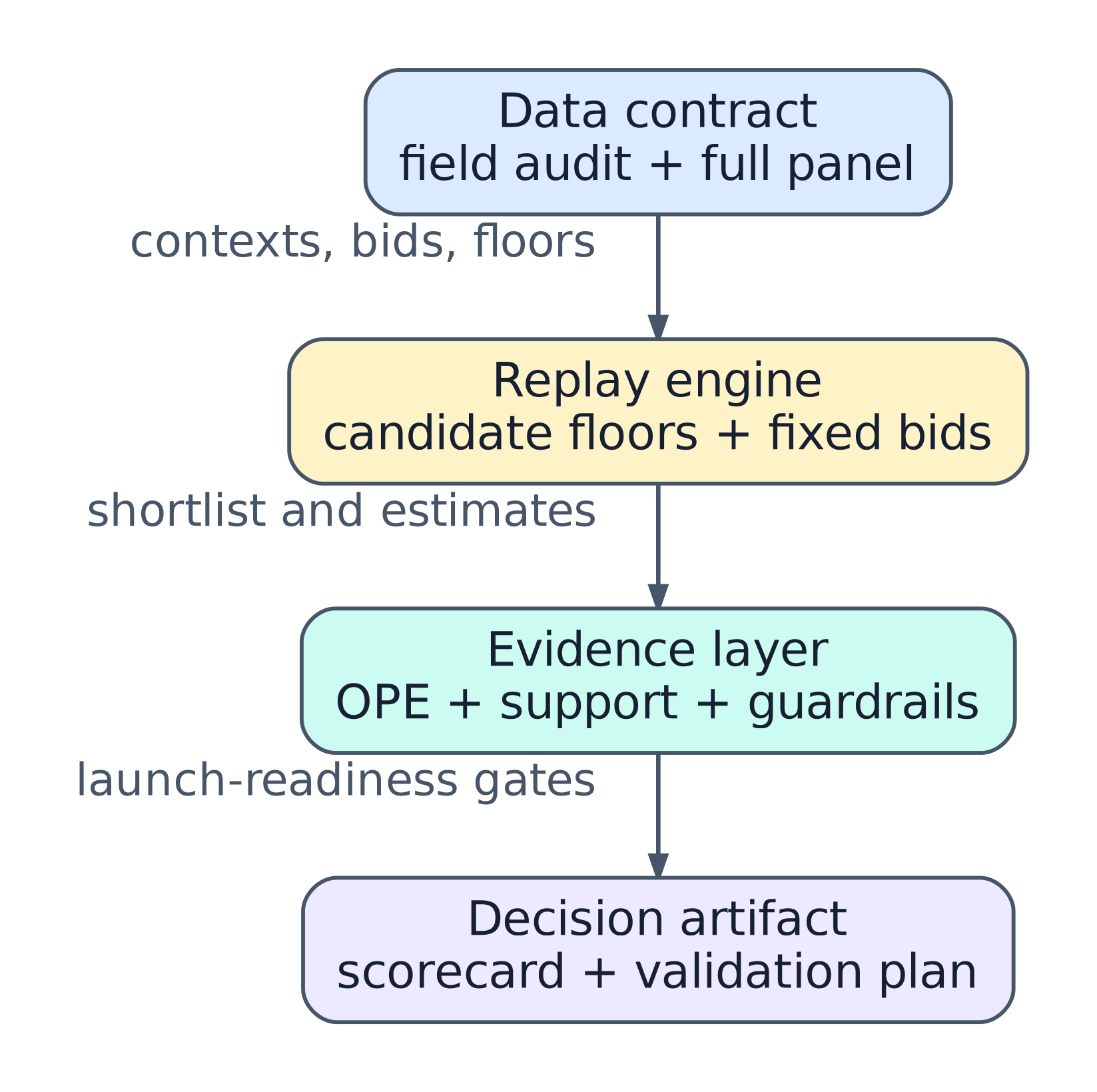}
    \vspace{-0.1em}
    {\\\small \textbf{(b)} Support-aware evaluation workflow.}
  \end{minipage}
  \caption{\small Marketplace interference and decision-support workflow for reserve/floor-policy evaluation. 
(a) \emph{Decision gap under interference:} Nodes represent auction opportunities linked through shared advertiser state (e.g., budgets and pacing). \emph{Solid arrows} denote the offline evaluation path used for policy selection, where effects are estimated from logged data via replay or OPE under a fixed marketplace state. \emph{Dashed arrows} denote the true evaluation path after deployment, where policies must be tested online and their effects propagate through shared state, altering future auctions via budget, pacing, and competition dynamics. These feedback effects are not captured in static logs, creating a gap between offline evidence and the live marketplace impact. 
(b) on the other hand shows the proposed \emph{Decision-support workflow.} In particular, logged auctions are transformed into replay outcomes, support-aware OPE diagnostics, guardrails, and sensitivity artifacts, which are combined into a launch-readiness recommendation (launch, validate, hold, or redesign).}
  \label{fig:interference_framework}
\end{figure}

Figure~\ref{fig:interference_framework}(b) presents the full workflow. The first layer is a measurement step that defines fields, estimands, policies, and assumptions. The second layer is empirical evaluation constructing the full panel, replay candidate floors, fit nuisance models, and estimate support-aware OPE diagnostics. The third layer is decision translation which combines point estimates, lower-tail estimates, guardrails, sensitivity checks, and validation feasibility into a launch-readiness decision artifact.

\subsection{Candidate policy class}

Let $\Pi=\{\pi_0,\pi_1,\ldots,\pi_{18}\}$ denote the finite set of candidate policies evaluated in this empirical study (see Table~\ref{tab:policyset} for the exhaustive list), where $\pi_0$ is the logged status quo. Each policy maps the observed auction context to a candidate floor,
\begin{equation}
  \pi_j:X_i\mapsto f_i^{\pi_j}, \qquad j=0,\ldots,18.
\end{equation}

The policy class is deliberately simple. Uniform policies scale all logged floors, absolute-increment policies add a fixed amount, minimum-floor policies impose quantile-based thresholds, and margin-gated policies intervene only when the observed bid--floor gap is sufficiently large. Throughout, quantiles (e.g., $q_{75}$) denote empirical percentiles of the logged floor distribution computed on the discovery panel.

A representative hybrid rule is
\begin{equation}
  f_i^{\pi}
  =
  \begin{cases}
    \max(f_i^0,q_{75}), & b_i-f_i^0\geq 100,\\
    f_i^0, & b_i-f_i^0<100,
  \end{cases}
\end{equation}
which corresponds to the \prioritypolicy{} selected later as the validation-priority candidate. This rule raises floors to a high (75th-percentile) level only when the observed bid--floor margin indicates sufficient headroom, and otherwise leaves the logged floor unchanged.

The policy class is not intended to exhaust all possible reserve-pricing algorithms. Its purpose is to construct auditable and interpretable candidates whose economic behavior can be inspected before considering more complex or adaptive pricing systems.

\subsection{Replay value and replay lift}

For each candidate policy, the static replay value is
\begin{equation}
  V_R(\pi) = \frac{1}{n}\sum_{i=1}^n
  D_i^0 \mathbf{1}\{b_i \geq f_i^\pi\}\max(p_i, f_i^\pi).
\end{equation}
Here $b_i$ denotes the observed bid (or winning-bid proxy) used to determine whether the auction would clear under the candidate floor, while $p_i$ denotes the realized payment. The replay estimand adjusts payments mechanically via $\max(p_i, f_i^\pi)$ rather than re-running the full auction. The corresponding replay lift relative to the logged baseline is
\begin{equation}
  \hat{\Delta}_R(\pi)
  =
  \frac{V_R(\pi)-V_R(\pi_0)}{V_R(\pi_0)}.
\end{equation}
For non-decreasing floor policies, this replay estimand has a clear mechanical interpretation stating that among the logged opportunities, which filled impressions would still clear and what they would pay if the candidate floor had been imposed while bids and participation stayed fixed. The method therefore uses replay as a screening layer, not as a live causal effect. A policy with high $\Delta_R(\pi)$ is eligible for deeper diagnostics, but it is not launch-ready until support, guardrails, response risk, and interference risk are assessed.

\subsection{Support-aware OPE layer}

The second layer asks whether the replay leader remains credible once the analysis treats counterfactual evidence as support-dependent. For each logged auction $i\in\mathcal{I}$, let $A_i$ denote the \emph{observed action} (e.g., the logged floor or its discretized representation), while $a$ denotes a generic action value in the action space. Let $e(a\mid X_i)$ denote the logging propensity (the probability that action $a$ was taken in context $X_i$), and let $m(X_i,a)$ be an outcome model. For a deterministic target policy $\pi$, a generic doubly robust (DR) score can be written as
\begin{equation}
  \psi_i^{DR}(\pi)
  =
  m(X_i,\pi(X_i))
  +
  \frac{\mathbf{1}\{A_i=\pi(X_i)\}}{e(\pi(X_i)\mid X_i)}
  \{Y_i-m(X_i,A_i)\}.
\end{equation}
The corresponding estimator averages these per-auction scores,
\[
{V}_{DR}(\pi)=\frac{1}{n}\sum_{i=1}^n\psi_i^{DR}(\pi).
\]
Using this, the OPE-based lift relative to baseline is defined as
\begin{equation}
  \hat{\Delta}_{DR}(\pi)
  =
  \frac{{V}_{DR}(\pi)-{V}_{DR}(\pi_0)}{{V}_{DR}(\pi_0)}.
\end{equation}
In production, this expression would require known or well-instrumented logging propensities. In the public-log setting studied here, the OPE layer is therefore interpreted as a support-aware diagnostic rather than as a definitive production causal estimate.

The methodology records support in three ways. First, it checks whether the target action $\pi(X_i)$ is observed (exactly or approximately) in the logged data. Second, it reports effective sample size (ESS),
\begin{equation}
  \mathrm{ESS}(\pi)
  =
  \frac{\left(\sum_i w_i^\pi\right)^2}{\sum_i (w_i^\pi)^2},
  \qquad
  w_i^\pi=\frac{\mathbf{1}\{A_i=\pi(X_i)\}}{e(\pi(X_i)\mid X_i)},
\end{equation}
which summarizes how concentrated the evaluation weights are across auctions. Third, it studies clipping sensitivity by replacing $w_i^\pi$ with $w_{i,c}^\pi=\min(w_i^\pi,c)$ over plausible caps $c$. A policy whose ranking survives these support checks receives stronger validation priority than a policy whose evidence depends on a few extreme weights or an arbitrary clipping choice.

% \subsection{Conservative ranking and guardrails}

% The ranking target is a conservative lower-tail lift, not the largest point estimate. Let $\hat{\Delta}(\pi)$ be an estimated lift and $\widehat{\mathrm{se}}\{\hat{\Delta}(\pi)\}$ its standard error. A one-sided lower-bound score is
% \begin{equation}
%   L_\alpha(\pi)
%   =
%   \hat{\Delta}(\pi)
%   -
%   z_{1-\alpha}\widehat{\mathrm{se}}\{\hat{\Delta}(\pi)\}.
% \end{equation}
% The validation-priority (VP) policy is then selected by
% \begin{equation}
%   \pi^{VP}\in\arg\max_{\pi\in\Pi} L_\alpha(\pi),
% \end{equation}
% subject to guardrail constraints. If $G_k(\pi)$ denotes guardrail $k$ and $\tau_k$ is the maximum tolerated deterioration, the feasible validation set is
% \begin{equation}
%   \Pi_{\mathrm{guard}}
%   =
%   \{\pi\in\Pi: G_k(\pi)\geq -\tau_k \ \text{for all guardrails } k\}.
% \end{equation}
% The empirical guardrails include retained-impression share, click and conversion proxies, advertiser-value proxies, daily stability, and segment-level harm. This structure keeps the methodology from becoming a single-metric revenue maximization exercise.

\begin{table}[t]
\centering
\caption{Guardrails used to screen reserve/floor policies before deeper evaluation.}
\label{tab:guardrails}
\begin{tabular}{p{0.22\linewidth} p{0.28\linewidth} p{0.18\linewidth} p{0.24\linewidth}}
\toprule
\textbf{Guardrail} & \textbf{Quantity Checked} & \textbf{Pass Rule} & \textbf{Purpose} \\
\midrule
Positive yield gain &
Percentage lift in yield per opportunity relative to the logged-floor baseline:
$\Delta Y_p / Y_0$ &
$\Delta Y_p / Y_0 \geq 0.005$ &
Removes policies whose replayed revenue gain is economically negligible or negative. \\

Aggregate fill retention &
Retained filled impressions under policy $p$ divided by observed filled impressions:
$I_p/I_0$ &
$I_p/I_0 \geq 0.98$ &
Prevents policies that increase yield mainly by dropping too much delivery volume. \\

Daily fill retention &
Minimum daily retained impression share:
$\min_t I_{p,t}/I_{0,t}$ &
$\min_t I_{p,t}/I_{0,t} \geq 0.98$ &
Rules out policies that look acceptable in aggregate but create severe delivery loss on at least one day. \\

Click retention &
Retained clicks under policy $p$ divided by baseline retained clicks:
$C_p/C_0$ &
$C_p/C_0 \geq 0.97$ &
Protects advertiser and member-engagement proxy quality from policies that remove high-click impressions. \\

Conversion retention &
Retained conversions under policy $p$ divided by baseline retained conversions:
$V_p/V_0$ &
$V_p/V_0 \geq 0.90$ &
Provides a conservative screen for deeper advertiser-value loss, recognizing that conversions are sparse. \\

Value-proxy retention &
Retained value proxy under policy $p$ relative to baseline:
$Q_p/Q_0$, where $Q=\text{clicks}+10\cdot\text{conversions}$ &
$Q_p/Q_0 \geq 0.97$ &
Combines click and conversion information into a single advertiser-value guardrail. \\

Yield stability &
Number of days with positive daily yield lift relative to the number of observed days &
Positive daily yield lift on every observed Season 2 day &
Rejects policies whose aggregate gain is driven by unstable or isolated daily wins. \\
\bottomrule
\end{tabular}
\end{table}

\subsection{Conservative ranking and guardrails}

The ranking target is a conservative lower-tail lift, not the largest point estimate. Let ${\Delta}(\pi)$ denote an estimated lift (e.g., replay or OPE-based lift relative to baseline) and $\widehat{\mathrm{se}}\{{\Delta}(\pi)\}$ its standard error. A one-sided lower-bound score is
\begin{equation}
  L_\alpha(\pi)
  =
  {\Delta}(\pi)
  -
  z_{1-\alpha}\widehat{\mathrm{se}}\{{\Delta}(\pi)\}.
\end{equation}
where $z_{1-\alpha}$ is the $(1-\alpha)$ quantile of the standard normal distribution. We use this lower-bound score to select the validation-priority (VP) policy as follows. Let $G_k(\pi)$ denote the change in guardrail metric $k$ under policy $\pi$ relative to baseline, and let $\tau_k$ denote the maximum tolerated deterioration. The feasible validation set is then defined as
\begin{equation}
  \Pi_{\mathrm{guard}}
  =
  \{\pi\in\Pi: G_k(\pi)\geq -\tau_k \ \text{for all guardrails } k\}.
\end{equation}

The empirical guardrails include retained-impression share (fill), click and conversion proxies, advertiser-value proxies, daily stability, and segment-level outcomes. Each guardrail is computed as a relative change from the logged baseline using the same replay or diagnostic layer used for the primary metric. Table~\ref{tab:guardrails} summarizes the launch-screening guardrails used after deterministic auction replay. A policy is marked eligible for deeper off-policy and sensitivity analysis only if all listed guardrails pass. This structure ensures that policies are not selected solely on revenue gains, but must also satisfy multi-sided marketplace constraints. Using the feasible validation set, the validation-priority (VP) policy is then selected by
\begin{equation}
  \pi^{VP}\in\arg\max_{\pi\in \Pi_{\mathrm{guard}}} L_\alpha(\pi),
\end{equation}

% \subsection{Out-of-time validation layer}

% When a later logged market window is available, the framework freezes the policy class and evaluates temporal transfer before converting evidence into a recommendation. Let $\mathcal{L}^{(2)}$ denote the discovery panel and $\mathcal{L}^{(3)}$ denote a later validation panel. The candidate set $\Pi$, floor quantiles, margin thresholds, and reader-facing policy definitions are learned or fixed before looking at $\mathcal{L}^{(3)}$. For each policy $\pi$, the validation replay lift is
% \begin{equation}
%   \Delta_R^{(3)}(\pi)
%   =
%   \frac{V_R^{(3)}(\pi)-V_R^{(3)}(\pi_0)}{V_R^{(3)}(\pi_0)}.
% \end{equation}
% The temporal-transfer gate is
% \begin{equation}
%   \begin{aligned}
%   T(\pi)=
%   \mathbf{1}\{&
%   \Delta_R^{(3)}(\pi)>0,\ 
%   \mathrm{rank}^{(3)}(\pi)\leq r_{\max},\\
%   &\mathrm{Retain}^{(3)}_m(\pi)\geq \tau_m
%   \ \text{for all monitored guardrails }m
%   \}.
%   \end{aligned}
% \end{equation}
% This is not an online causal validation gate. It is an out-of-time replay gate: it asks whether the offline recommendation is stable when market composition changes and the analyst is not allowed to retune the rule. Passing $T(\pi)$ strengthens validation priority; it does not remove the need for shadow logging, real propensities, or an interference-aware online experiment.

\subsection{Out-of-time validation layer}

When a later logged market window is available, the framework freezes the policy class and evaluates temporal transfer before converting evidence into a recommendation. Let $\mathcal{L}^{(2)}$ denote the discovery panel and $\mathcal{L}^{(3)}$ denote a later validation panel. The candidate set $\Pi$, floor quantiles, margin thresholds, and reader-facing policy definitions are fixed before looking at $\mathcal{L}^{(3)}$, so that no retuning is performed on the validation window.

For each policy $\pi$, the validation replay lift is
\begin{equation}
  \Delta_R^{(3)}(\pi)
  =
  \frac{V_R^{(3)}(\pi)-V_R^{(3)}(\pi_0)}{V_R^{(3)}(\pi_0)}.
\end{equation}

The temporal-transfer gate is
\begin{equation}
  \begin{aligned}
  T(\pi)=
  \mathbf{1}\{&
  \Delta_R^{(3)}(\pi)>0,\ 
  \mathrm{rank}^{(3)}(\pi)\leq r_{\max},\\
  &\mathrm{Retain}^{(3)}_m(\pi)\geq \tau_m
  \ \text{for all monitored guardrails }m
  \},
  \end{aligned}
\end{equation}
where $\mathrm{rank}^{(3)}(\pi)$ denotes the policy's rank under validation replay lift among all candidates, and $\mathrm{Retain}^{(3)}_m(\pi)$ denotes the retained level of guardrail metric $m$ under replay in the validation window (e.g., retained impressions or value proxy relative to baseline). This gate therefore checks three conditions: (i) the policy remains beneficial out of time, (ii) it remains among the top candidates without retuning, and (iii) it does not violate guardrail constraints in the new market window. 

This is not an online causal validation gate. It is an out-of-time replay gate and asks whether the offline recommendation is stable when market composition changes and the analyst is not allowed to adjust the policy. Passing $T(\pi)$ strengthens validation priority, but does not remove the need for shadow logging, real propensities, or an interference-aware online experiment.

\subsection{Launch-readiness decision rule}

The final step of the framework maps evidence into an operational action. Rather than combining all metrics into a single score, the framework evaluates whether different types of evidence support specific claims. Let $\mathbf{G}(\pi)=(G_1(\pi),\ldots,G_K(\pi))$ denote the vector of guardrail outcomes, and define the decision map
\begin{equation}
\label{dpi}
d(\pi)=\mathcal{D}\{V_R(\pi),\ \Delta_R(\pi),\ T(\pi),\ L_\alpha(\pi),\ \mathrm{ESS}(\pi),\ \mathbf{G}(\pi),\ B(\pi),\ I(\pi)\}.
\end{equation}

Each input corresponds to a different type of evidence. Replay value $V_R(\pi)$ and replay lift $\Delta_R(\pi)$ measure mechanical performance under fixed bids. The temporal-transfer gate $T(\pi)$ checks whether the policy remains effective out of time under a frozen specification. The lower-bound score $L_\alpha(\pi)$ and effective sample size $\mathrm{ESS}(\pi)$ capture statistical reliability under limited support. The guardrail vector $\mathbf{G}(\pi)$ enforces marketplace constraints such as retained impressions, value proxies, and segment-level effects. 

The remaining terms capture risks that cannot be resolved from static logs alone. The response robustness $B(\pi)$ measures how much adverse marketplace response (e.g., bidder adaptation or pacing changes) the policy can tolerate while still outperforming the baseline. The interference indicator $I(\pi)$ represents whether this remaining uncertainty has been resolved through an interference-aware validation design, such as shadow logging with observed propensities and a switchback experiment.

To make the decision rule explicit, each type of evidence is converted into a binary gate. For a policy $\pi$, define:
\begin{itemize}
\item $R(\pi)$: replay upside (whether $\Delta_R(\pi)>0$),
\item $T(\pi)$: out-of-time transfer (stability across time),
\item $S(\pi)$: support adequacy (sufficient ESS and overlap),
\item $C(\pi)$: conservative evidence (positive lower-bound lift),
\item $H(\pi)$: guardrail passage (no constraint violations),
\item $B(\pi)$: response robustness (sufficient break-even margin),
\item $I(\pi)$: interference-resolved evidence (validated online).
\end{itemize}

These gates are combined into two summary indicators:
\[
Q_{-I}(\pi)=R(\pi)\,T(\pi)\,S(\pi)\,C(\pi)\,H(\pi)\,B(\pi), 
\qquad
Q(\pi)=Q_{-I}(\pi)\,I(\pi).
\]

Here $Q_{-I}(\pi)=1$ means that the policy passes all offline and sensitivity-based checks, while $Q(\pi)=1$ additionally requires that interference and response uncertainty have been resolved through online validation.

The resulting decision rule is
\begin{equation}
\label{dpi2}
d(\pi)=
\begin{cases}
\text{launch}, & Q(\pi)=1,\\
\text{validate online}, & Q_{-I}(\pi)=1,\ I(\pi)=0,\\
\text{hold}, & \text{evidence is mixed but diagnosable},\\
\text{redesign}, & \text{core measurement or guardrails fail}.
\end{cases}
\end{equation}

This structure makes the decision rule intentionally conservative. Offline replay, OPE diagnostics, and sensitivity analyses can establish that a policy is \emph{validation-ready}, meaning it is a strong candidate for experimentation. However, direct launch requires passing an additional gate involving evidence that the policy performs well under real marketplace dynamics, including bidder response and interference. The separation between $Q_{-I}(\pi)$ and $Q(\pi)$ therefore encodes the central distinction of the framework which states that strong offline evidence is sufficient to justify validation, but not sufficient to justify deployment.

\begin{algorithm}[!htbp]
\caption{Support-aware launch-readiness evaluation}
\label{alg:launch-readiness}
\footnotesize
\begin{algorithmic}[1]
\Require Logged auction panel $\mathcal{L}$; policy class $\Pi$; guardrail thresholds $\{\tau_k\}_{k=1}^{K}$; support thresholds; response-sensitivity thresholds; validation-design constraints.
\Ensure Launch-readiness artifact containing a recommended policy, supported claims, blocking assumptions, and next action.
\State Construct auditable non-decreasing candidate floor policies $\pi\in\Pi$ and record their operational rules.
\For{each candidate policy $\pi\in\Pi$}
  \State Estimate replay value $V_R(\pi)$ and replay lift $\Delta_R(\pi)$.
  \State Compute basic retained-impression and marketplace-quality guardrails.
\EndFor
\State Remove dominated candidates whose replay lift, retained-impression share, or basic guardrails are inferior to available alternatives.
\If{a later logged validation panel is available}
  \State Freeze $\Pi$ and all policy thresholds; replay each surviving policy on the validation panel.
  \State Record temporal-transfer gate $T(\pi)$ using validation-period lift, rank, and guardrail retention.
\Else
  \State Set $T(\pi)=1$ as a non-blocking placeholder and record that out-of-time validation evidence is unavailable.
\EndIf
\State Fit nuisance and support models needed for the diagnostic OPE layer.
\For{each surviving policy $\pi$}
  \State Estimate support-aware scores, lower-tail value $L_\alpha(\pi)$, effective sample size $\mathrm{ESS}(\pi)$, and clipping sensitivity.
  \State Apply guardrail constraints $\mathbf{G}(\pi)$ for fill, quality proxies, segment harm, and daily stability.
  \State Compute break-even response robustness $B(\pi)$.
  \If{$R(\pi)T(\pi)S(\pi)C(\pi)H(\pi)B(\pi)I(\pi)=1$}
    \State Assign $d(\pi)=\text{launch}$.
  \ElsIf{$R(\pi)T(\pi)S(\pi)C(\pi)H(\pi)B(\pi)=1$ and $I(\pi)=0$}
    \State Assign $d(\pi)=\text{validate online}$.
  \ElsIf{evidence is mixed but diagnosable}
    \State Assign $d(\pi)=\text{hold}$.
  \Else
    \State Assign $d(\pi)=\text{redesign}$.
  \EndIf
\EndFor
\State Select the highest-priority non-dominated policy under the action map $d(\pi)=\mathcal{D}\{\cdot\}$ from (\ref{dpi}).
\State \Return launch-readiness artifact and recommended next experiment.
\end{algorithmic}
\end{algorithm}

To enforce clarity in presentation, we also represent the entire idea of the paper as Algorithm~\ref{alg:launch-readiness}. The individual estimators and plots are known ingredients, but the algorithm specifies how they are sequenced, what each is allowed to claim, and how the final action is chosen when identification remains incomplete. The empirical study therefore evaluates the algorithm against simplified decision rules that use only replay, only OPE, only guardrails, or only out-of-time replay. This ablation tests whether integration changes the decision itself rather than only adding documentation around a conventional policy ranking. The next section states the formal empirical results that justify these methodological gates.

\begin{table}[!t]
\centering
\begin{threeparttable}
\caption{Core decision-support gates.}
\label{tab:gates}
\papertablesize
\renewcommand{\arraystretch}{1.12}
\begin{tabularx}{\linewidth}{lXX}
\toprule
Gate & Evidence used & Decision role \\
\midrule
Replay evidence & Full-panel mechanical yield and retained-impression share & Screens candidate policies \\
Out-of-time validation & Frozen-policy replay on later market window & Tests whether the offline recommendation transfers without retuning \\
Support and overlap & Effective sample size, exact floor support, clipping sensitivity & Determines whether OPE estimates are credible \\
Guardrails & Fill, advertiser value proxy, click/conversion proxy, segment harm & Prevents single-metric revenue maximization \\
Sensitivity & Support collapse, break-even bidder response, placebo checks & Downgrades claims that rely on fragile assumptions \\
Interference-aware validation & Shadow logging, switchback detectability, stop rules & Converts offline priority into an online test plan \\
\bottomrule
\end{tabularx}
\end{threeparttable}
\end{table}

\section{Empirical Study}

The empirical study is organized as a decision story rather than as a sequence of disconnected model outputs. It begins with the logged auction environment, asks whether reserve/floor changes are economically plausible, constructs auditable candidate policies, then progressively downgrades or strengthens the evidence using stability, support, uncertainty, sensitivity, and guardrail checks. The goal is not only to find a policy with high replay value. The goal is to decide whether the available evidence supports direct launch, online validation, redesign, or no action.

\subsection{Dataset and validation windows}

The empirical analysis uses two non-overlapping iPinYou training windows. Season two is the discovery and main analysis window as it is used to build the full auction panel, define the replay frontier, estimate nuisance models, rank candidate policies, and construct the decision artifact. Season three is reserved for external replay validation and hence the policy definitions, quantile thresholds, and margin gates are frozen before season-three replay. This split is important because the season-three results are not another tuning pass, rather they test whether the selected policy transfers to a later market window. It should be noted that we create a 60\% training split inside the Season 2 development panel, followed by 20\% for validation, and the final 20\% for testing. This split is used whenever the pipeline estimates or evaluates learned nuisance and outcome models. Precisely, the LightGBM fill, pay-price, click, conversion, and value-proxy models are trained on the \texttt{train} slice and reported on held-out data; the calibration diagnostics use held-out predictions; the direct-method and doubly robust off-policy evaluation components use model predictions in a way intended to avoid evaluating a model only on rows used to fit it; and the cross-fitted doubly robust analysis further strengthens this by refitting outcome models across folds on sampled Season 2 rows. However, it should be noted that the split is not used to limit the deterministic auction replay as reserve/floor policy replay evaluates all configured Season 2 opportunity shards, because replay is a mechanical counterfactual calculation based on logged bid prices, logged floors, fills, pay prices, clicks, and conversions rather than a learned predictive evaluation.

\begin{table}[!t]
\centering
\begin{threeparttable}
\caption{Distribution of the discovery and external-validation data windows.}
\label{tab:datawindows}
\papertablesize
\setlength{\tabcolsep}{2.0pt}
\renewcommand{\arraystretch}{1.12}
\begin{tabularx}{\linewidth}{
  >{\raggedright\arraybackslash}p{0.12\linewidth}
  >{\raggedright\arraybackslash}p{0.18\linewidth}
  >{\raggedleft\arraybackslash}p{0.06\linewidth}
  >{\raggedleft\arraybackslash}p{0.14\linewidth}
  >{\raggedleft\arraybackslash}p{0.14\linewidth}
  >{\raggedleft\arraybackslash}p{0.07\linewidth}
  >{\raggedleft\arraybackslash}p{0.07\linewidth}
  >{\raggedleft\arraybackslash}p{0.08\linewidth}}
\toprule
Window & Role in study & Dates & Bid opportunities & Filled impressions & Fill rate & Clicks & Conver\-sions \\
\midrule
Season two & Discovery and main empirical panel & 7 & 53{,}289{,}330 & 12{,}190{,}344 & 22.9\% & 8{,}729 & 391 \\
Season three & Out-of-time frozen-policy validation & 9 & 10{,}566{,}743 & 3{,}132{,}311 & 29.6\% & 2{,}691 & 526 \\
\bottomrule
\end{tabularx}
\begin{tablenotes}
\papertablesize
\item Season three contains about 19.8\% as many bid opportunities and 25.7\% as many filled impressions as season two. Conversion counts are read from the available conversion logs; later season-three dates without conversion files are treated as zero observed conversions in the validation artifact.
\end{tablenotes}
\end{threeparttable}
\end{table}

Table~\ref{tab:datawindows} here shows that the holdout window is smaller in traffic volume but still large enough for a meaningful replay validation. Season two contains 53.3 million bid opportunities across seven days, while season three contains 10.6 million opportunities across nine days. The average daily volume therefore falls from about 7.6 million opportunities per day in season two to about 1.2 million in season three. The fill rate is higher in season three, and the observed conversion count is not lower despite the smaller opportunity set. This distribution shift is useful for validation because if the leading policy remained strong only because season two was unusually high-volume, the season-three replay would expose that fragility.

\subsection{Policy set and season-two discovery results}

The pipeline constructs a full season-two panel from iPinYou-style RTB logs. Candidate policies include minimum-floor rules, margin-aware increments, and hybrid rules that raise floors only when logged bid-floor margins are sufficiently large. The policy family is intentionally auditable, or in other words it is simple enough for operational teams to reason about and rich enough to reveal yield and support tradeoffs. This design choice is important for a DSS setting. A black-box floor policy could potentially improve a metric, but it would be harder for stakeholders to understand why the recommendation changed and harder to translate into a controlled marketplace experiment.

Table~\ref{tab:policyset} lists the full candidate set used in the replay comparison. The table uses reader-facing policy names rather than raw code identifiers, while the reproducibility artifacts preserve the implementation identifiers. This makes the comparison easier to read without hiding the operational rule being tested.

\begin{table}[!t]
\centering
\begin{threeparttable}
\caption{Candidate reserve/floor policies compared in the empirical study.}
\label{tab:policyset}
\papertablesize
\renewcommand{\arraystretch}{1.08}
\begin{tabularx}{\linewidth}{
  >{\raggedright\arraybackslash}p{0.09\linewidth}
  >{\raggedright\arraybackslash}p{0.22\linewidth}
  >{\raggedright\arraybackslash}p{0.28\linewidth}
  >{\raggedright\arraybackslash}X}
\toprule
Policy number & Policy family & Reader-facing policy & Operational rule \\
\midrule
P0 & Baseline & Logged status quo & Use the logged floor exactly as observed. \\
P1 & Uniform percentage raise & Uniform +5\% & Raise every logged floor by 5\%. \\
P2 & Uniform percentage raise & Uniform +10\% & Raise every logged floor by 10\%. \\
P3 & Uniform percentage raise & Uniform +15\% & Raise every logged floor by 15\%. \\
P4 & Uniform percentage raise & Uniform +20\% & Raise every logged floor by 20\%. \\
P5 & Uniform percentage raise & Uniform +30\% & Raise every logged floor by 30\%. \\
P6 & Absolute increment & Add 5 to all floors & Add 5 price units to every logged floor. \\
P7 & Absolute increment & Add 10 to all floors & Add 10 price units to every logged floor. \\
P8 & Absolute increment & Add 20 to all floors & Add 20 price units to every logged floor. \\
P9 & Minimum positive floor & Positive-floor q25 minimum & Raise positive logged floors below q25 to q25. \\
P10 & Minimum positive floor & Positive-floor q50 minimum & Raise positive logged floors below q50 to q50. \\
P11 & Minimum positive floor & Positive-floor q75 minimum & Raise positive logged floors below q75 to q75. \\
P12 & Minimum all floor & All-floor q25 minimum & Raise all logged floors below q25 to q25, including zero floors. \\
P13 & Minimum all floor & All-floor q50 minimum & Raise all logged floors below q50 to q50, including zero floors. \\
P14 & Margin-aware increment & Gap-25 add 5 & Add 5 when the bid-floor gap is at least 25. \\
P15 & Margin-aware increment & Gap-50 add 10 & Add 10 when the bid-floor gap is at least 50. \\
P16 & Margin-aware increment & Gap-100 add 20 & Add 20 when the bid-floor gap is at least 100. \\
P17 & Hybrid minimum-margin rule & Q50 Margin-Gated Floor & Raise to q50 only when the bid-floor gap is at least 50. \\
P18 & Hybrid minimum-margin rule & Q75 Margin-Gated Floor & Raise to q75 only when the bid-floor gap is at least 100. \\
\bottomrule
\end{tabularx}
\end{threeparttable}
\end{table}

\begin{figure}[!t]
  \centering
  \includegraphics[width=0.68\linewidth,height=0.40\textheight,keepaspectratio]{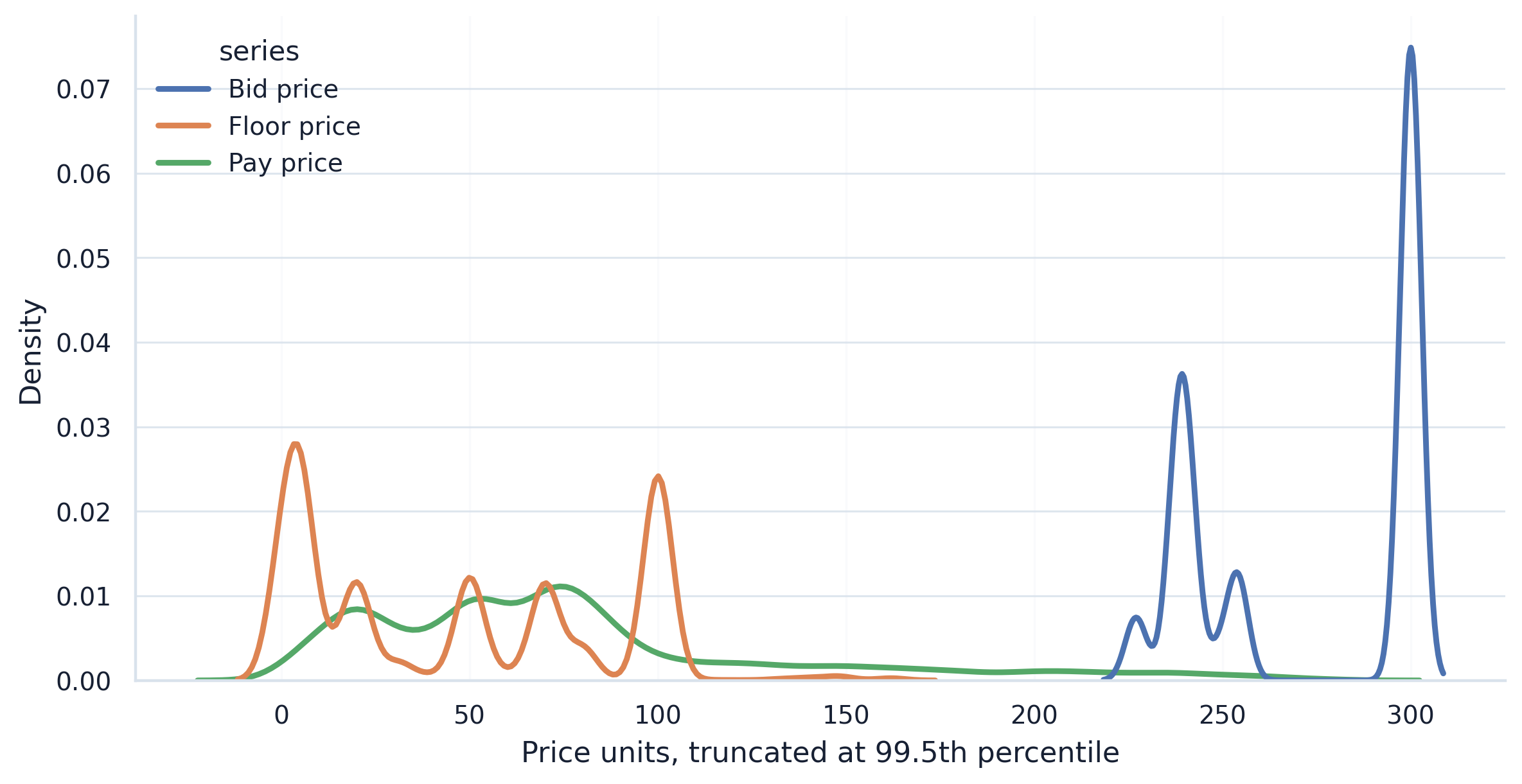}
  \caption{Sample bid, payment, and floor distributions from the data audit. The heavy-tailed price landscape motivates support diagnostics before reserve/floor-policy replay.}
  \label{fig:prices}
\end{figure}

\begin{figure}[!t]
  \centering
  \includegraphics[width=0.68\linewidth,height=0.40\textheight,keepaspectratio]{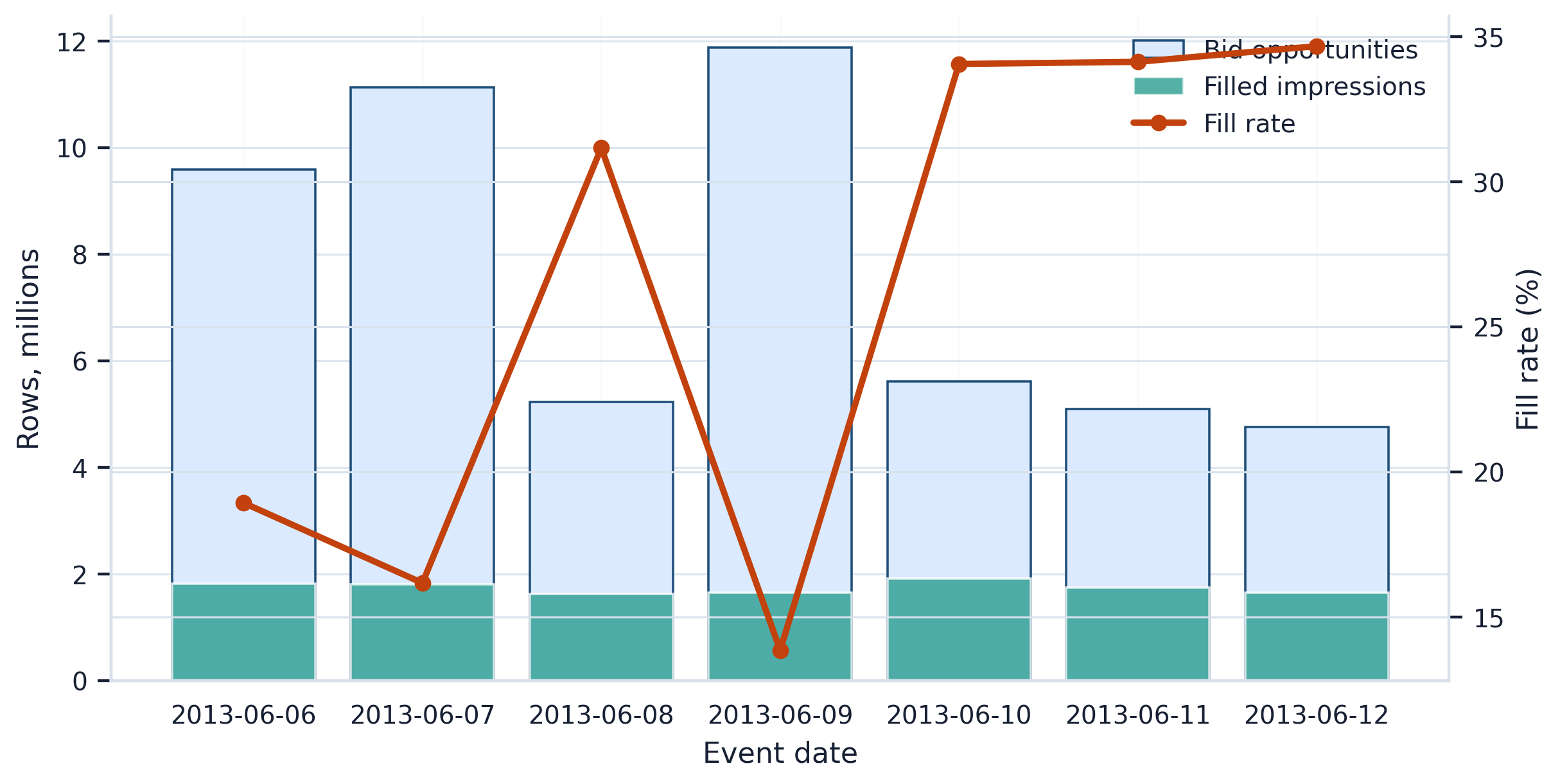}
  \caption{\small Outcome density by day in the full season-two panel. The main empirical results are based on the processed full-panel artifact rather than only on audit samples.}
  \label{fig:outcomedensity}
\end{figure}

\begin{figure}[!t]
  \centering
  \includegraphics[width=0.99\linewidth,height=0.40\textheight,keepaspectratio]{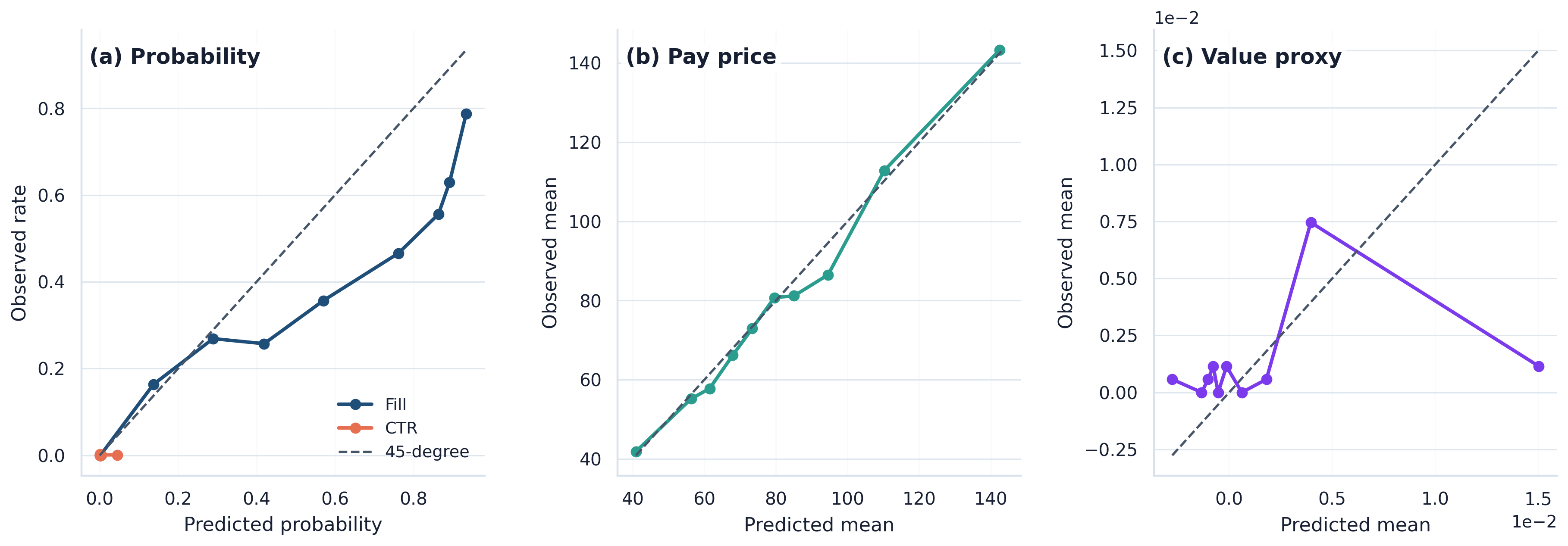}
\caption{\small Nuisance-model diagnostics for the model-assisted decision layer. Panel (a) reports held-out probability calibration for the LightGBM classification models, estimated with Python's \texttt{lightgbm} implementation through \texttt{scikit-learn} pipelines: the fill model estimates $\Pr(\mathrm{filled}=1\mid x)$ and the click-through-rate model estimates $\Pr(\mathrm{clicked}=1\mid \mathrm{filled}=1,x)$. Panels (b) and (c) report binned held-out regression calibration for LightGBM regression models predicting pay price and the constructed value proxy, respectively. In each panel, points compare the mean model prediction within a prediction bin to the corresponding observed mean outcome; the dashed line denotes perfect calibration. The value-proxy panel is visibly weaker because the target is constructed from sparse click and conversion events; this limits the weight placed on value-model evidence and motivates treating value preservation as a guardrail rather than as a standalone launch criterion.}
\label{fig:nuisance}

\end{figure}

Figures~\ref{fig:prices}--\ref{fig:nuisance} summarize the empirical foundation before the policy experiments. The price landscape motivates reserve/floor interventions, the outcome-density figure verifies that the full panel covers the observed season-two days, and the nuisance-model diagnostics show that the OPE layer is not operating without predictive signal.

Figure~\ref{fig:prices} establishes the first empirical fact that the Season 2 auction environment has enough price dispersion for floor policy to matter. The figure is generated from a 600{,}000-row sample drawn from the full seven-day Season 2 opportunity panel, so it is a scalable audit view rather than a one-day snapshot. If bids, payments, and existing floors were tightly concentrated, most candidate floor changes would either do nothing or remove nearly all fill. Instead, the observed distributions show a wide and asymmetric price landscape that many opportunities have substantial bid-floor margin, while others sit close to the logged floor or realized payment. This makes the policy problem nontrivial. Higher floors can plausibly increase yield on high-margin opportunities, but the same intervention can destroy delivery or value on low-margin opportunities. The figure therefore motivates the paper's selective policy class, which raises floors only when the logged opportunity appears to have sufficient margin, and it also motivates the downstream guardrails and support checks used before any launch recommendation.

Figure~\ref{fig:outcomedensity} moves the story from price dispersion to panel coverage and verifies that the empirical study is built on a full configured Season 2 opportunity panel with observable daily scale, fills, and outcome density. Daily outcome density matters because the later replay and guardrail analyses aggregate across days while also checking whether candidate policies remain stable day by day. If evidence were driven by only one or two unusually active days, or if the panel construction produced uneven delivery coverage, the replay frontier would be a weak basis for experiment planning. The observed full-panel construction gives the later policy comparisons a more credible operational base and motivates reporting both aggregate and daily guardrails.

Figure~\ref{fig:nuisance} evaluates the nuisance-model layer used by the model-assisted parts of the decision support system. The figure appears before the policy results because replay and OPE use different kinds of evidence. Auction replay is mechanical in the sense that given a logged bid and a proposed floor, we can check whether the bid would still clear and what payment would be retained. OPE and doubly robust diagnostics require additional learned outcome components. The nuisance models are trained and evaluated on a reproducible chronological sample drawn from the full Season 2 panel, rather than on the entire 53.3 million-row panel. This keeps model fitting tractable while preserving the broader Season 2 setting from which the sample is drawn. Fig.~\ref{fig:nuisance} therefore asks whether those learned components behave plausibly on held-out data (test split) before they are used downstream.

Panel~(a) reports probability calibration for two classification nuisance models. The first is the fill model, which estimates
\[
\widehat p_{\mathrm{fill}}(x_i)
=
\Pr(\mathrm{filled}_i=1 \mid x_i),
\]
the probability that bid opportunity \(i\) becomes a filled impression given pre-auction features \(x_i\). The second is the click-through-rate model, which estimates
\[
\widehat p_{\mathrm{click}}(x_i)
=
\Pr(\mathrm{clicked}_i=1 \mid \mathrm{filled}_i=1, x_i),
\]
the probability that a filled impression receives a click. The horizontal axis gives the average predicted probability within a prediction bin, and the vertical axis gives the realized event rate in that same bin. If a bin has mean predicted fill probability of 0.30, for example, good calibration means that roughly 30\% of rows in that bin are actually filled. The diagonal line represents perfect calibration. The fill model is especially important because floor changes mainly affect the delivery channel; the click model is noisier because clicks are much rarer, so its evidence is treated more cautiously.

Panel~(b) gives the corresponding binned diagnostic for the pay-price regression model. This model estimates the expected payment conditional on the pre-auction features and the filled-impression population. Held-out rows are sorted into bins by predicted pay price. Each point compares the mean predicted payment in a bin with the mean observed payment in that bin. This diagnostic matters because revenue effects are not determined only by fill retention. Conditional on an impression being retained, the payment distribution determines how much yield the marketplace keeps or gains under a candidate floor.

Panel~(c) repeats the same regression-calibration check for the value-proxy model. The value proxy summarizes sparse downstream response outcomes into a single score used to stress-test whether candidate floor policies preserve advertiser and user-response value, rather than only increasing short-run platform yield. Because these downstream events are sparse, this panel is interpreted as a guardrail diagnostic rather than as a precise prediction exercise.

Taken together, the three panels show that the paper treats nuisance models as inspected inputs, not as black-box decision makers. Good calibration strengthens confidence in direct-method and doubly robust evidence. Weak or noisy calibration does not invalidate replay, but it reduces the weight placed on model-assisted counterfactual scores and increases the importance of guardrails, sensitivity analysis, Season 3 validation, and eventual online testing. With the nuisance layer audited, we now move to the substantive policy question about which reserve/floor policies are worth considering at all?
\begin{figure}[!t]
  \centering  \includegraphics[width=0.58\linewidth,height=0.40\textheight,keepaspectratio]{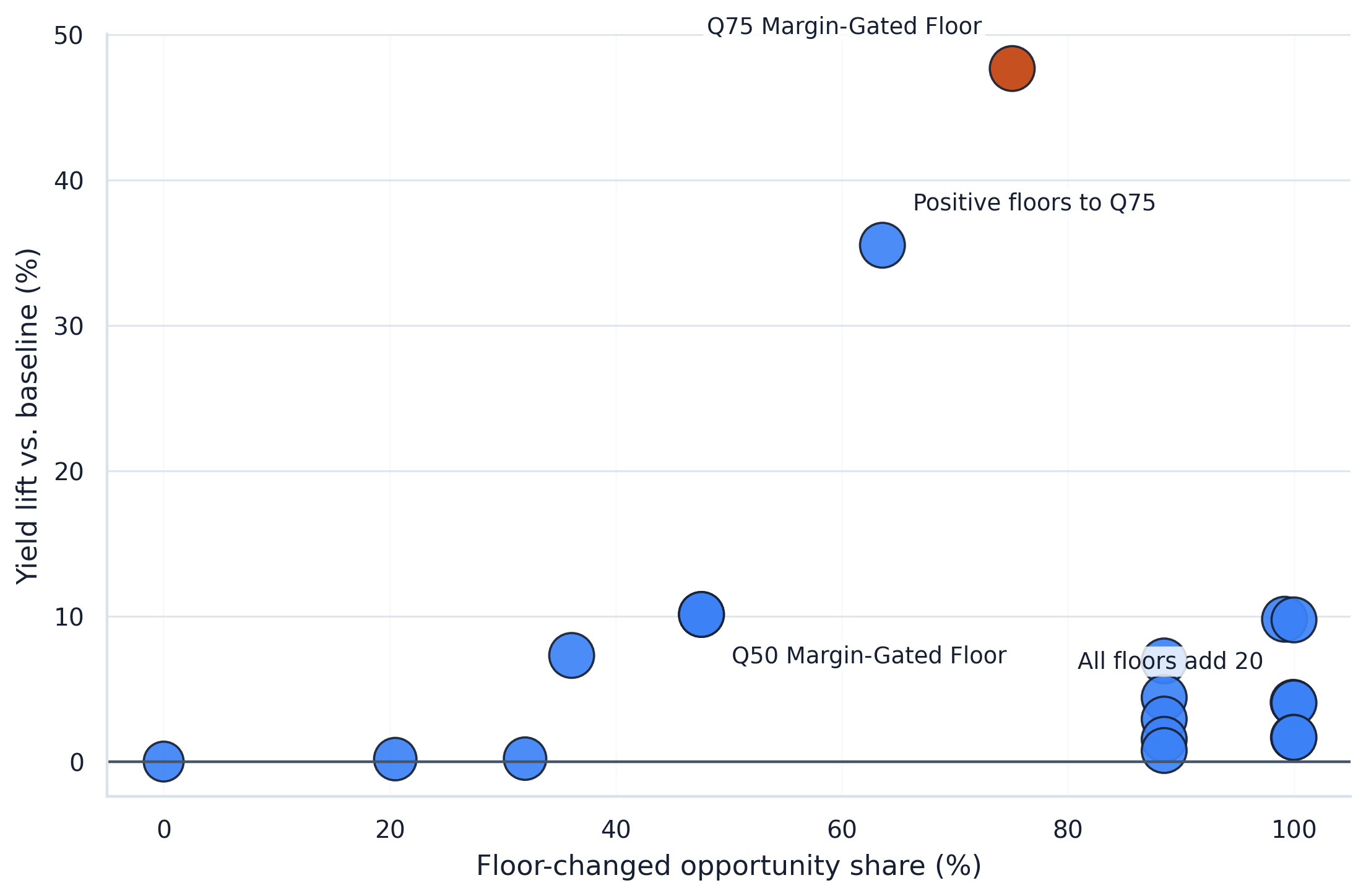}
  \caption{\small Reserve/floor policy replay frontier on the Season 2 panel. Each point is one candidate policy from the reserve-policy catalog. The horizontal axis reports the share of bid opportunities whose floor would change under the policy, and the vertical axis reports replayed yield lift relative to the logged-floor baseline. Marker size reflects the number of replay guardrails passed, and the priority policy, Q75 Margin-Gated Floor, is highlighted in red. The figure is a screening diagnostic: it identifies policies with attractive mechanical yield upside under fixed logged bids, not launch-ready causal treatment effects.}
\label{fig:frontier}
\end{figure}
Figure~\ref{fig:frontier} reports the Season 2 replay tradeoff frontier for the reserve/floor policy catalog. This is the first point where the empirical analysis produces a policy-relevant finding. Each point represents a candidate floor rule evaluated under the same mechanical replay contract. In essence, logged bids are held fixed, and then the candidate floor is applied to each opportunity, and replayed yield is computed from the retained impressions. The horizontal axis measures intervention breadth, namely the share of opportunities whose floor would change. The vertical axis measures replayed yield lift relative to the logged-floor baseline. Marker size summarizes how many guardrails the policy passes, so the figure displays both upside and operational risk in the same view. The strongest replay candidate is the \prioritypolicy{}, a hybrid rule that raises floors to the q75 level only when the bid-floor gap is at least 100. In full-panel replay, this policy produces a 47.7\% yield lift while retaining the logged filled impressions under the replay contract. The frontier also shows why the problem is a tradeoff rather than a one-dimensional maximization. Some policies create yield lift by changing a large share of auctions; such broad interventions may be harder to defend operationally and may be more exposed to bidder response. The leading hybrid rule is attractive because it is selective and raises floors where logged bid margins suggest room to move, while leaving thin-margin opportunities unchanged. Thus, the frontier is not used to declare launch readiness. It is used to identify a credible shortlist for deeper OPE, support, sensitivity, and validation analysis.

\begin{figure}[!t]
  \centering
  \includegraphics[width=0.78\linewidth,height=0.40\textheight,keepaspectratio]{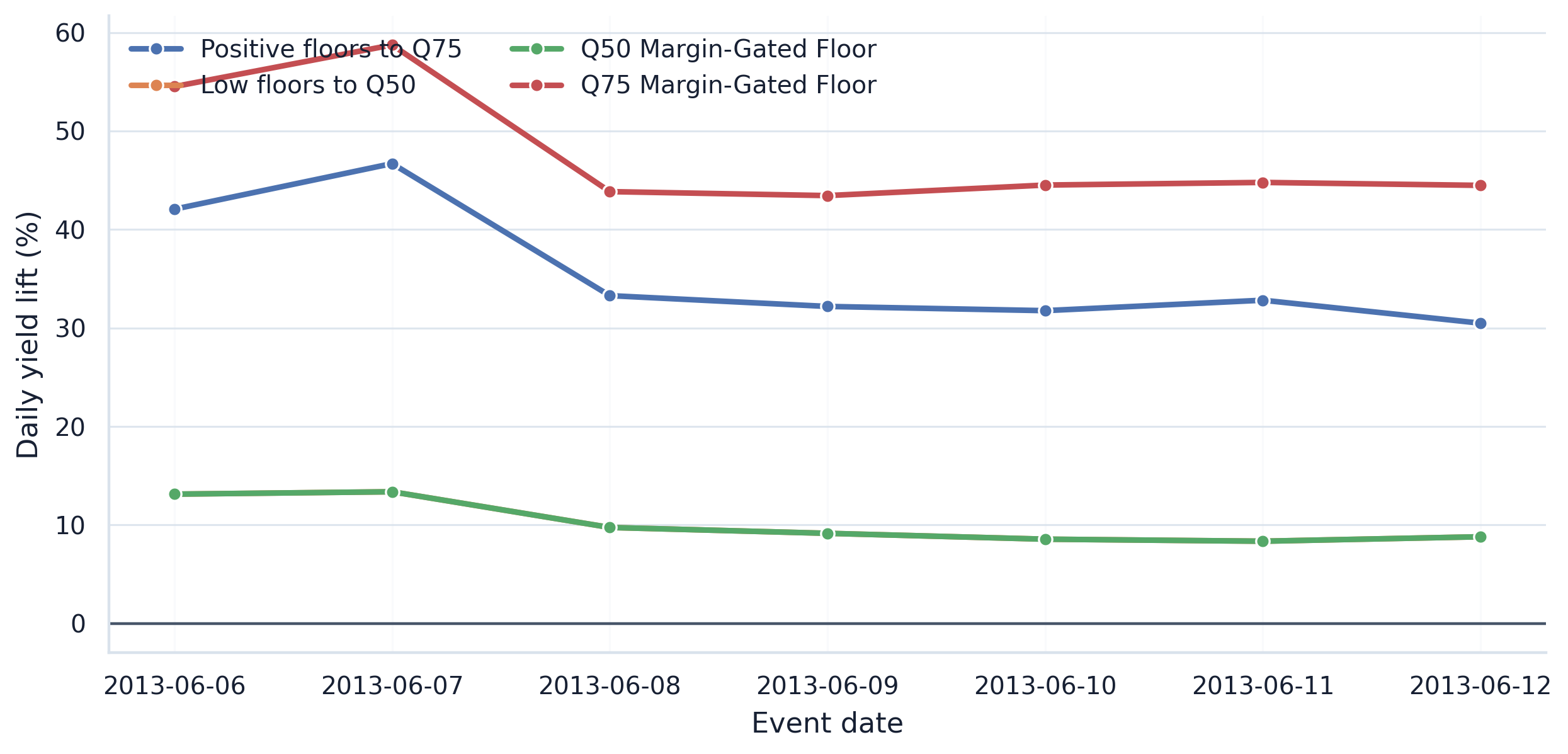}
  \caption{\small Daily stability of shortlisted reserve/floor policies on the Season 2 panel. Each line reports daily replayed yield lift relative to the logged-floor baseline for one shortlisted policy. The horizontal zero line marks no lift over the baseline. The purpose of the figure is to check whether aggregate replay gains are repeatable across days rather than driven by a single unusual traffic or auction condition. Stable positive daily lift strengthens the case for deeper OPE, sensitivity, and validation analysis, but does not by itself establish launch readiness.}
\label{fig:daily}

\end{figure}

Figure~\ref{fig:daily} tests whether the replay frontier survives a basic temporal-stability check. A policy that looks attractive only because of one unusual day is a weak candidate for a costly marketplace experiment, even if its aggregate replay lift is large. The daily-stability view therefore takes the shortlisted policies from the replay and guardrail screen and plots their daily yield lift relative to the logged-floor baseline. The figure shows that the leading policies have positive replay lift throughout the Season 2 window rather than relying on a single traffic spike or isolated market condition. This matters because the replay frontier in Fig.~\ref{fig:frontier} is an aggregate view; it can hide instability across days. Daily stability does not make the estimate causal, and it does not solve bidder-response or pacing concerns. It does, however, strengthen the case that the leading policy is not merely an artifact of one anomalous slice of the data. With a stable replay leader identified, the analysis can move from mechanical screening to the harder questions involving whether the policy has adequate support for OPE, whether its ranking is robust to uncertainty and heterogeneity, and what validation design is needed before launch. Before moving from deterministic replay to model-assisted OPE, we apply an explicit guardrail screen to the replayed policies. The purpose of this step is to prevent the analysis from treating aggregate yield lift as sufficient evidence. A floor policy can increase replayed yield while still creating unacceptable delivery loss, value-proxy loss, or temporal instability. We therefore evaluate each policy against seven pre-specified screens: (i) positive yield lift, (ii) aggregate fill retention, (iii) daily fill retention, (iv) click retention, (v) conversion retention, (vi) value-proxy retention, and (vii) daily yield stability. More details on these guardrails are provided in table \ref{tab:guardrails}.

\begin{figure}[!t]
\centering
\includegraphics[width=0.75\linewidth]{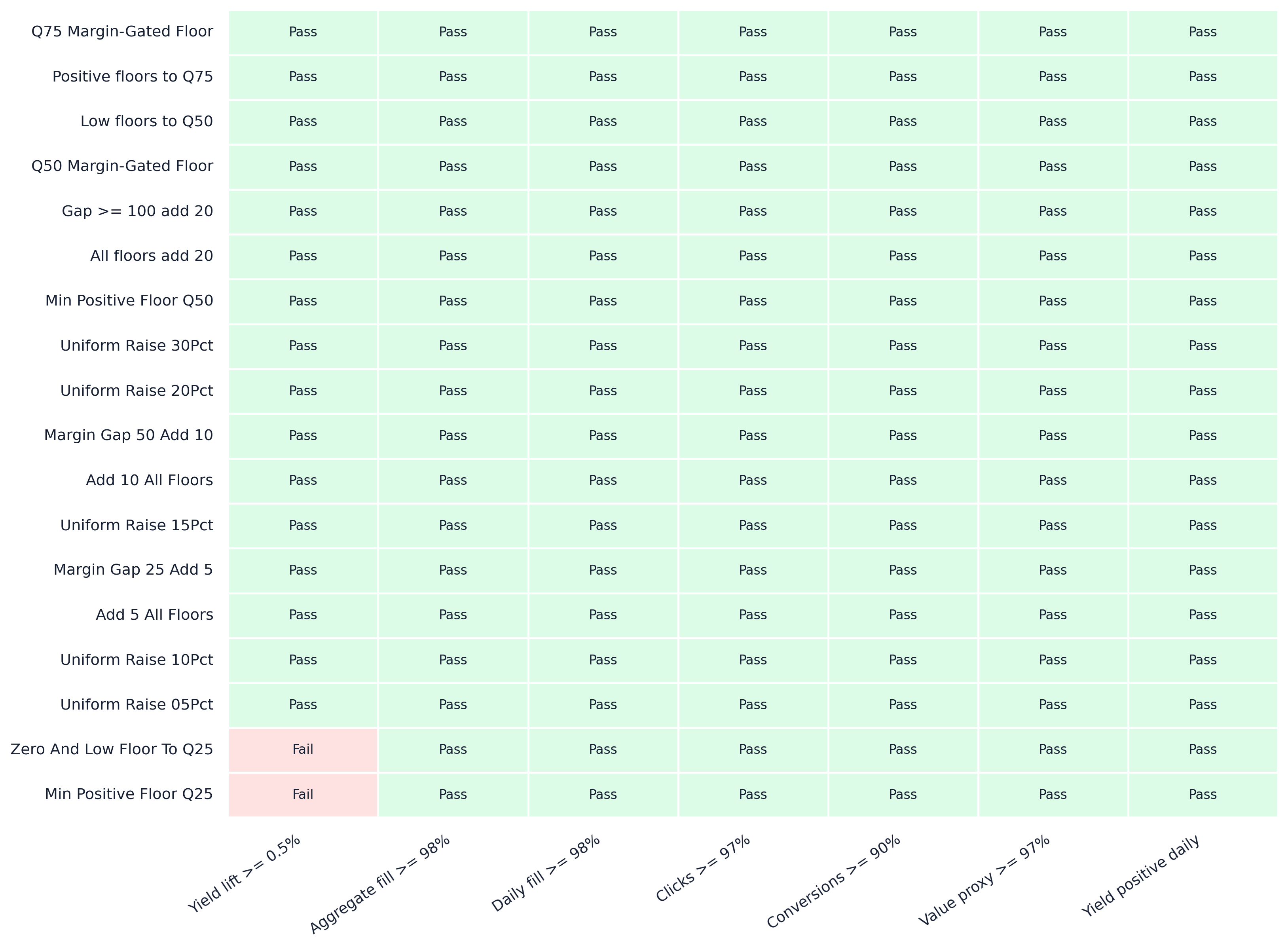}
\caption{\small Replay guardrail matrix for reserve-policy candidates. Rows are candidate floor policies and columns are pre-specified replay guardrails. A policy passes the full screen only if it has at least 0.5\% replay yield lift, at least 98\% aggregate retained impressions, at least 98\% minimum daily retained impressions, at least 97\% retained clicks, at least 90\% retained conversions, at least 97\% retained value proxy, and positive yield lift on every observed Season 2 day. The matrix makes the candidate handoff auditable before OPE: policies are not advanced solely because they have high replay yield.}
\label{fig:guardrail_matrix}
\end{figure}

Figure~\ref{fig:guardrail_matrix} reports the pass/fail status of each candidate policy under these replay guardrails. This figure shows whether a policy is operationally safe enough to justify deeper analysis. In particular, the screen separates policies that merely increase replayed yield from policies that preserve the basic marketplace quantities needed for a credible validation candidate.

This guardrail stage is the first narrowing point in the DSS. Policies that pass the screen are not declared launch-ready; they are only considered eligible for more expensive diagnostics, including support-aware OPE, conservative lower-tail ranking, heterogeneity analysis, robustness checks, and Season 3 validation. This ordering is important for the paper's logic: OPE is not used to rescue policies that already fail basic replay safety checks, and replay yield is not allowed to dominate the decision system without delivery and value safeguards.

% \begin{table}[!t]
% \centering
% \begin{threeparttable}
% \caption{Priority policy recommendation.}
% \label{tab:recommendation}
% \papertablesize
% \renewcommand{\arraystretch}{1.12}
% \begin{tabularx}{\linewidth}{lX}
% \toprule
% Decision item & Value \\
% \midrule
% Priority policy & \prioritypolicy{} \\
% Replay yield lift & 47.7\% \\
% Season-three external replay validation & Rank 1; 43.9\% yield lift; 100.0\% impression and value-proxy retention \\
% Conservative cross-fitted DR lower-tail lift & 45.8\% \\
% Weighted evidence score & 4.375 \\
% Break-even adverse marketplace-response loss & 32.3\% \\
% Direct launch ready & No \\
% Recommended action & Shadow logging followed by exchange-hour switchback \\
% \bottomrule
% \end{tabularx}
% \end{threeparttable}
% \end{table}

% Table~\ref{tab:recommendation} is the empirical hinge of the paper. It deliberately contains two facts that would be easy to separate in a conventional analysis: the priority policy has large offline upside, and the evidence is still not launch-ready. The contribution is the rule that keeps both facts visible at the same time. The replay lift and conservative DR lower-tail lift justify spending validation capacity on the policy; the missing real propensities, unobserved bidder response, and interference pathways block a direct launch claim. This is why the paper's output is a decision recommendation rather than only a policy ranking.

\subsection{Support-aware off-policy diagnostics}

The replay frontier identifies policies with attractive fixed-bid mechanical upside, but replay alone does not answer whether the policy ranking is robust to model-assisted logged-data evidence. We therefore add an off-policy-evaluation (OPE) diagnostic layer. Because the public iPinYou logs do not contain randomized assignment probabilities for the candidate floor policies, these estimators are not treated as production-valid launch estimators. Instead, the reproduction pipeline uses a simulated known-propensity logger over the replay shortlist. This allows us to audit estimator agreement, support, weight behavior, and lower-tail uncertainty under transparent assumptions.

The purpose of this section is therefore narrower than causal launch validation. It asks whether the priority policy remains credible when the replay evidence is passed through support-aware and model-assisted diagnostics. If the priority policy only looked favorable under mechanical replay, or if its OPE estimates depended on extreme weights, then the decision system would downgrade it before any online validation recommendation.

\begin{figure}[!t]
  \centering
  %\widepaperfig{08_ope_estimator_comparison.png}
  \includegraphics[width=0.78\linewidth]{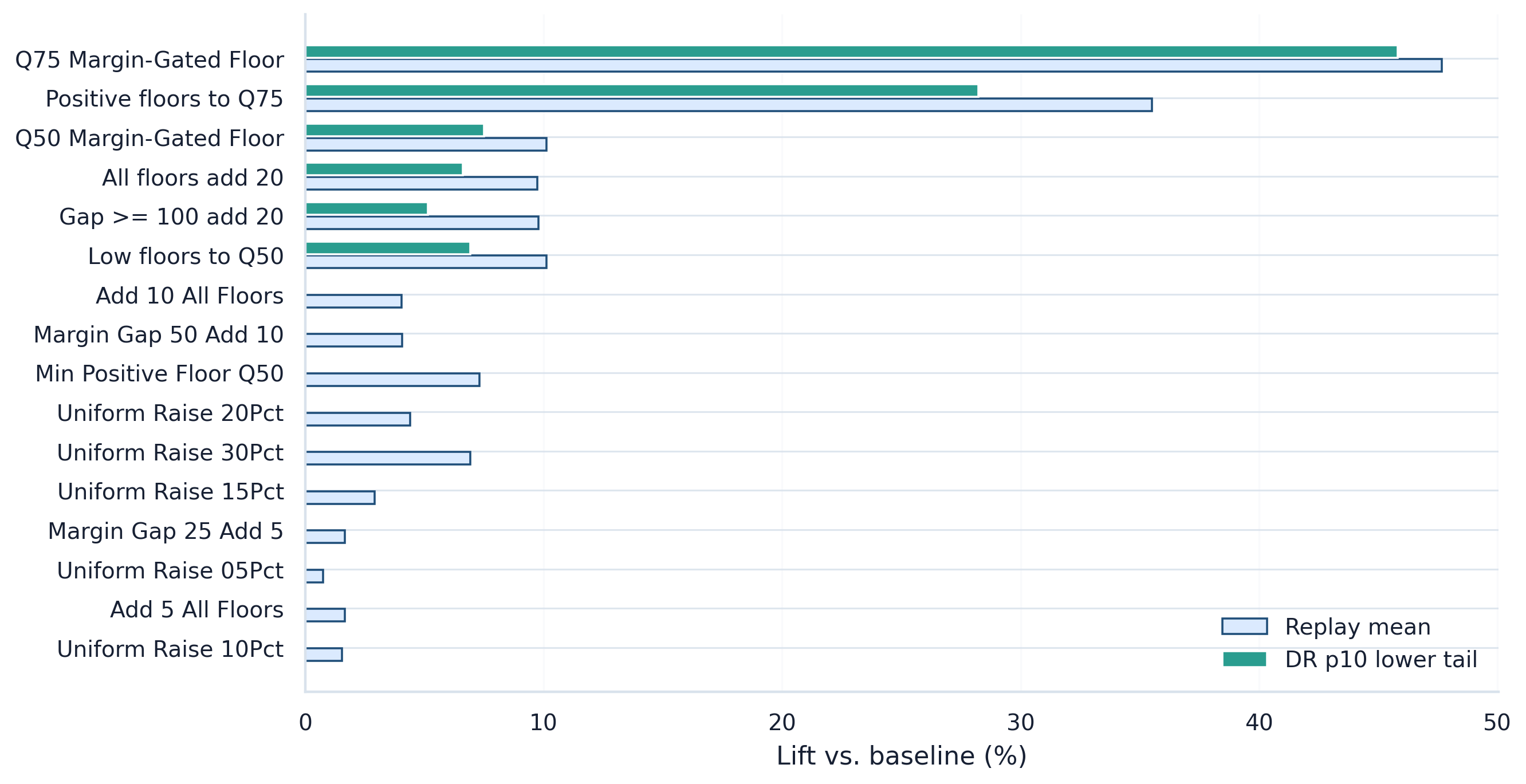}
  \caption{\small Estimator comparison for shortlisted reserve/floor policies. The figure compares replay mean lift with conservative lower-tail cross-fitted doubly robust (DR) lift under the simulated known-propensity logger. Agreement between replay and lower-tail DR diagnostics supports validation priority, while disagreement would indicate estimator or support fragility.}
  \label{fig:opecomparison}
\end{figure}

Figure~\ref{fig:opecomparison} compares the replay mean with the conservative lower-tail doubly robust diagnostic. The goal is to detect whether the priority policy is an artifact of a single estimator family. Replay answers a fixed-bid mechanical question, while the DR diagnostic combines an outcome model with inverse-propensity correction under the simulated logger. Agreement between these views is stronger evidence than either view alone. The important point is not that OPE proves launch readiness; it does not. Rather, the estimator comparison shows that the priority policy remains a credible validation candidate after moving from aggregate replay to model-assisted lower-tail evidence.

\begin{figure}[!t]
  \centering
  %\widepaperfig{08_weight_diagnostics.png}
  \includegraphics[width=0.78\linewidth]{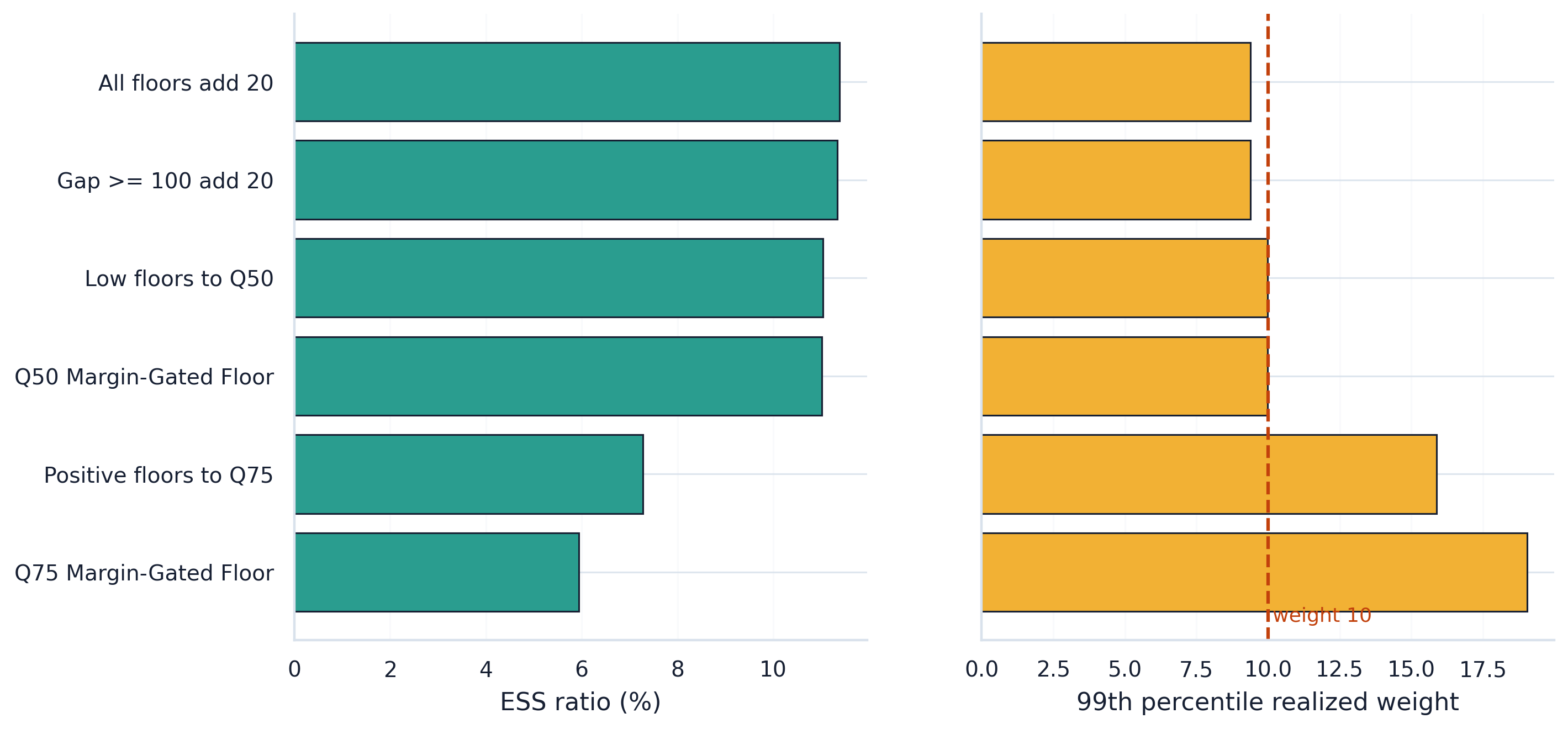}
  \caption{\small Importance-weight support diagnostics under the simulated known-propensity logger. The left panel reports effective sample size (ESS) as a share of the evaluation sample. The right panel reports the 99th percentile realized importance weight, with the vertical reference line marking weight 10. Weight support is treated as a decision gate: weak support reduces the credibility of OPE evidence and prevents direct launch claims.}
  \label{fig:weights}
\end{figure}

Figure~\ref{fig:weights} explains how much trust to place in the OPE layer. Importance weights reveal whether a target policy is evaluated on broad support or on a small number of influential observations. A policy can have an attractive mean estimate while being supported by only a thin part of the logged-data distribution. In that case, the OPE estimate is fragile even if the point estimate is positive. The effective-sample-size diagnostic and the upper-tail weight diagnostic therefore function as evidence gates rather than technical afterthoughts. Poor support does not invalidate replay, but it prevents model-assisted estimates from being interpreted as sufficient launch evidence.

\begin{figure}[!t]
  \centering
  %\paperfig{08_conservative_lower_bound_ranking.png}
  \includegraphics[width=0.78\linewidth]{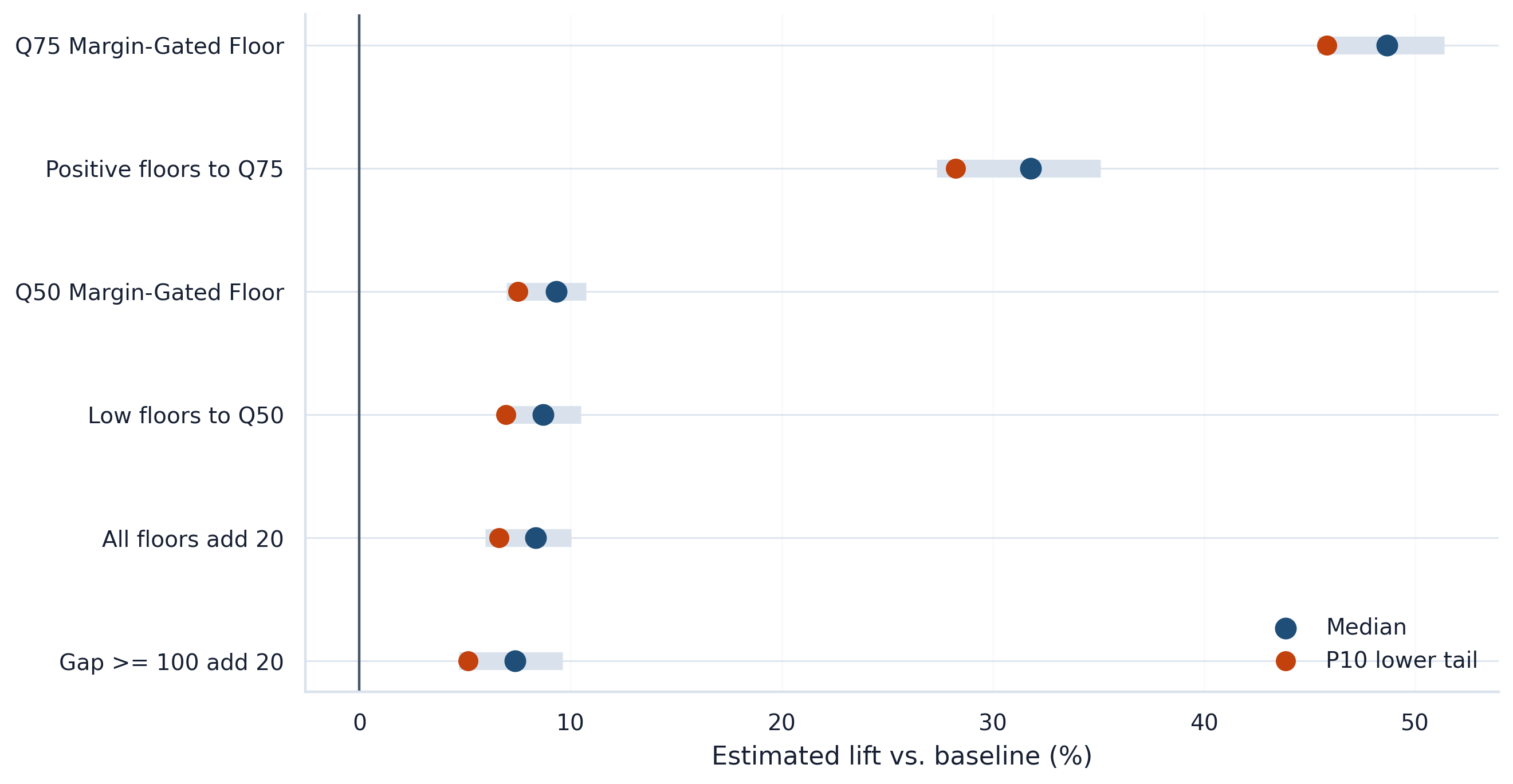}
  \caption{\small Conservative lower-bound ranking from cross-fitted DR diagnostics. Points show lower-tail and median bootstrap lift estimates for candidate policies. The ranking uses lower-tail evidence rather than mean lift alone, aligning the statistical diagnostic with the managerial question of which policy deserves scarce online validation capacity.}
  \label{fig:conservative}
\end{figure}

Figure~\ref{fig:conservative} converts the OPE diagnostics into a conservative validation-priority ranking. Instead of ranking policies only by their average estimated lift, the decision system uses lower-tail cross-fitted DR evidence. This is important because the operational decision is asymmetric as the platform does not need to launch the most optimistic offline estimate, but it does need to avoid spending experimentation capacity on a policy whose downside-adjusted evidence collapses. The priority policy remains attractive under this lower-tail criterion, which strengthens the case for online validation. At the same time, the figure reinforces the paper's central caution. Lower-tail OPE improves the evidence beyond replay-only screening, but it does not resolve live bidder response, interference, or the absence of real production propensities.

Together, Figures~\ref{fig:opecomparison}--\ref{fig:conservative} move the analysis from replay screening to support-aware validation priority. The priority policy is not advanced because it has the largest replay lift alone. It is validation worthy because replay, model-assisted diagnostics, support checks, and conservative lower-tail ranking all point in the same direction under clearly stated assumptions. The next step is to ask whether this favorable average evidence hides segment-level downside risk across exchanges, regions, support clusters, or time buckets. We therefore inspect segment-level heterogeneity for the priority policy before moving to broader sensitivity analysis. This step asks whether the favorable lower-tail OPE evidence is concentrated in only a few parts of the market, or whether it is visible across large operational segments.

\begin{figure}[!t]
  \centering
  %\widepaperfig{08_segment_heterogeneity.png}
  \includegraphics[width=0.78\linewidth]{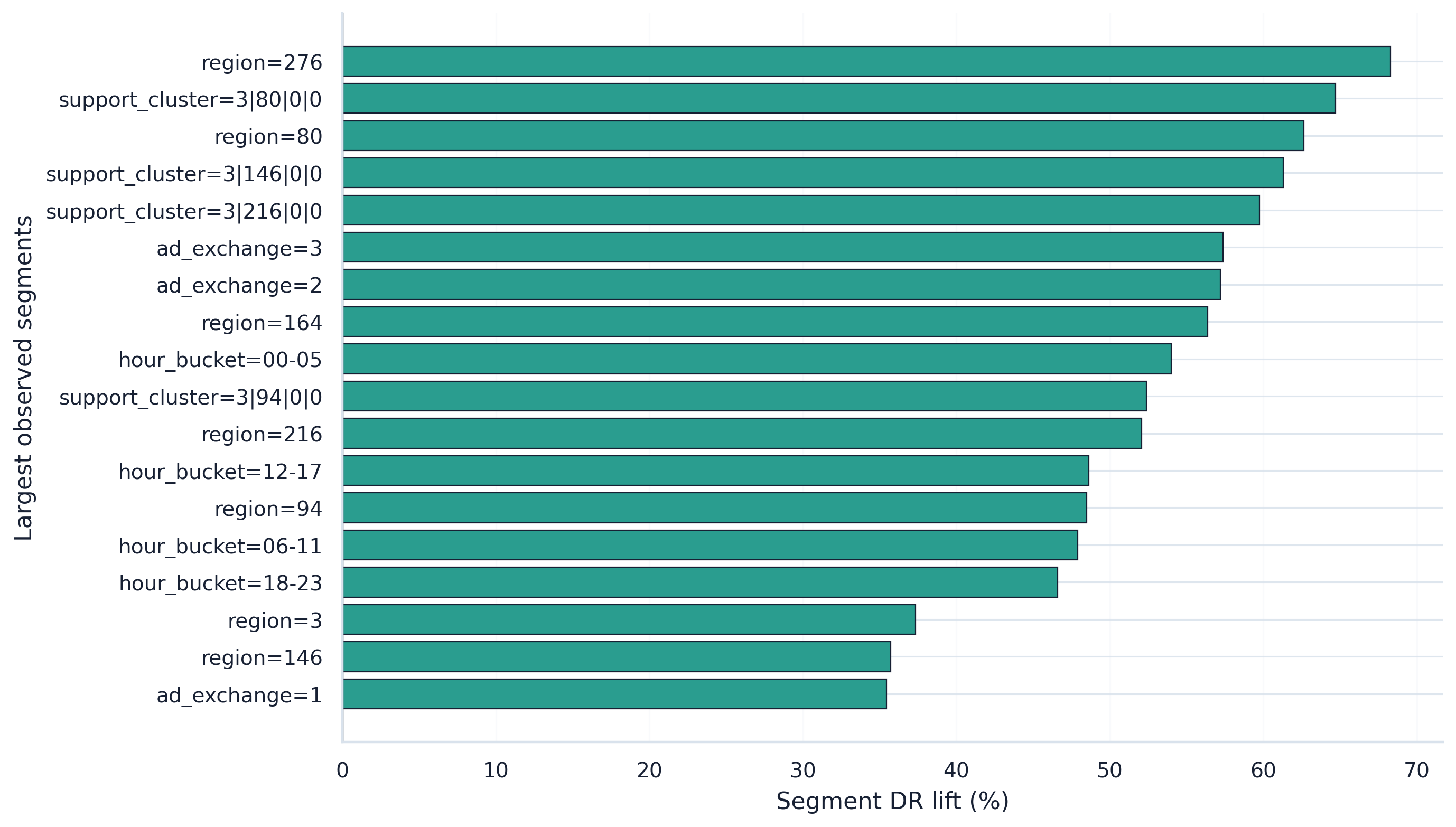}
  \caption{\small Segment-level heterogeneity for the priority reserve/floor policy. Bars report cross-fitted doubly robust (DR) lift estimates for the largest observed segment cells, including exchange, region, support-cluster, and time-bucket slices. Positive bars indicate segments where the priority policy improves estimated yield relative to the logged-floor baseline; negative bars would indicate segment-level downside. The figure checks whether favorable average OPE evidence hides obvious large-segment harm.}
  \label{fig:heterogeneity}
\end{figure}

Figure~\ref{fig:heterogeneity} shows that the priority policy's favorable evidence is not only an aggregate artifact. The largest observed segment cells generally retain positive DR lift, suggesting that the policy's average gain is not driven by one narrow market slice. This matters because the replay and OPE rankings are policy-level summaries. Without a segment view, a policy could appear attractive while exposing important submarkets to downside risk.

The heterogeneity diagnostic is still not a launch claim. Segment estimates inherit the limitations of the logged-data setting: they do not observe live bidder response, budget pacing changes, or interference under the new floor rule. Instead, the figure functions as a risk screen. If large segments showed negative lift, the decision system would need to narrow the eligible traffic, redesign the policy, or add segment-specific guardrails before validation. Since the priority policy does not exhibit obvious large-segment harm in this diagnostic, the analysis can proceed to the remaining unobserved threats: marketplace response, support collapse, and out-of-time validation.

\subsection{Robustness to Marketplace Response and Support Contraction}
\label{subsec:robustness}

The preceding analyses identify \prioritypolicy{} as the leading offline candidate under replay, doubly robust off-policy evaluation, and lower-tail ranking. However, a replay or OPE estimate is not yet a live-launch treatment effect. Two remaining threats are especially important in a reserve-price marketplace. First, bidders and budget-allocation systems may respond when floors change. Second, the OPE evidence may weaken if the effective support of the target policy is poorer than the observed diagnostic suggests. We therefore run two deterministic robustness experiments before making the validation recommendation.

The first experiment studies adverse marketplace response. Let $Y_0$ denote the logged-floor baseline yield per opportunity, and let $Y_{\pi}$ denote the replay yield per opportunity for candidate policy $\pi$. For an adverse response-loss share $\rho \in [0,0.60]$, the adjusted lift is defined as
\begin{equation}
\Delta_{\pi}^{\mathrm{resp}}(\rho)
=
\frac{Y_{\pi}(1-\rho)}{Y_0} - 1.
\label{eq:response_sensitivity}
\end{equation}
This calculation does not claim to forecast bidder behavior. Instead, it asks how much yield loss from bidder shading, budget reallocation, pacing response, or demand substitution would be required before the replay advantage is erased. The break-even response-loss share is
\begin{equation}
\rho_{\pi}^{\star}
=
1 - \frac{Y_0}{Y_{\pi}}
=
\frac{\Delta_{\pi}(0)}{1+\Delta_{\pi}(0)}.
\label{eq:breakeven_response}
\end{equation}
where $\Delta_{\pi}(0) = Y_{\pi}/Y_0 - 1$. For \prioritypolicy{}, the replay lift is $47.7\%$, implying a break-even adverse response-loss share of $32.3\%$.

\begin{figure}[!t]
\centering
\includegraphics[width=0.68\linewidth]{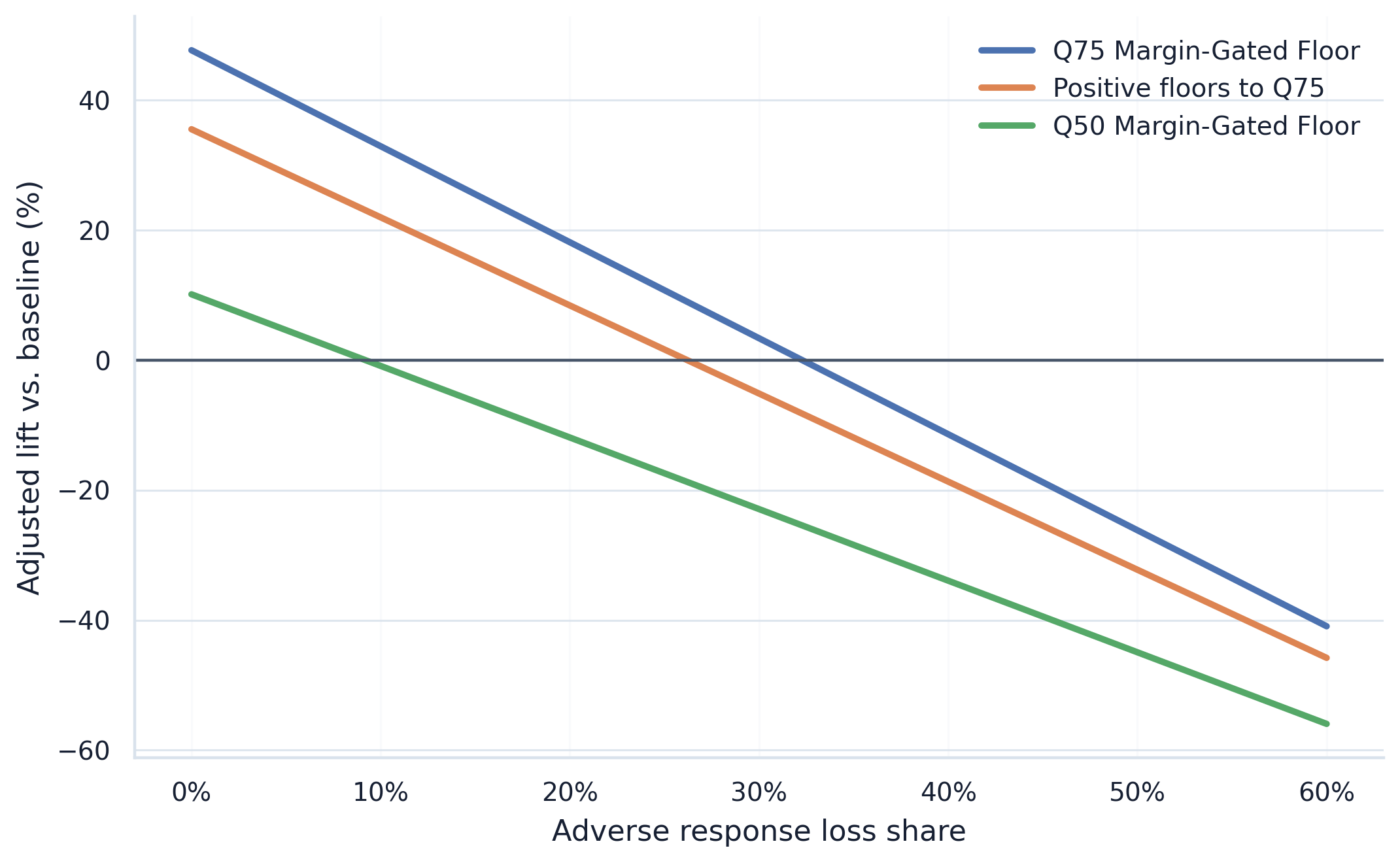}
\caption{\small Marketplace-response sensitivity. The figure applies an adverse response-loss share $\rho$ to each shortlisted policy's replay yield per opportunity and recomputes lift relative to the logged-floor baseline using Eq.~\eqref{eq:response_sensitivity}. The underlying artifact evaluates six shortlisted policies over 31 response-loss values from 0\% to 60\%; the plotted view shows the priority policy and two interpretable comparison policies for readability. \prioritypolicy{} remains positive until an approximately 32.3\% adverse response loss, indicating substantial offline upside but not proving launch readiness.}
\label{fig:response_sensitivity}
\end{figure}

Figure~\ref{fig:response_sensitivity} shows that the priority policy has materially more response-loss tolerance than the simpler comparison rules. This is important because the replay frontier alone could overstate launch confidence in assuming replay keeps participation and bids fixed. The response curve makes that limitation explicit. A policy with high replay lift is valuable only if the implied improvement is not fragile to plausible marketplace adjustment. The priority policy survives a large adverse-response stress test, which supports allocating validation capacity to it, but the same analysis also explains why the paper stops short of recommending direct launch.

The second experiment studies support contraction in the OPE layer. Let $m_{\pi}$ be the cross-fitted DR median lift for policy $\pi$, and let $\ell_{\pi}$ be its observed lower-tail lift, here the bootstrap p10 estimate. If effective support declines by a scale factor $s \in \{1.00,0.75,0.50,0.25,0.10,0.05\}$, the support-adjusted lower-tail lift is defined as
\begin{equation}
\ell_{\pi}^{\mathrm{sup}}(s)
=
m_{\pi}
-
\frac{m_{\pi}-\ell_{\pi}}{\sqrt{s}}.
\label{eq:support_sensitivity}
\end{equation}
The square-root inflation reflects the usual sampling-rate relationship between effective sample size and standard-error scale, which states that when effective support falls, lower-tail uncertainty expands approximately at rate $1/\sqrt{s}$. This is not a new estimator; it is a stress test that keeps the observed DR median and lower-tail gap fixed while asking how conservative the lower bound becomes under poorer overlap.

\begin{figure}[!t]
\centering
\includegraphics[width=0.68\linewidth]{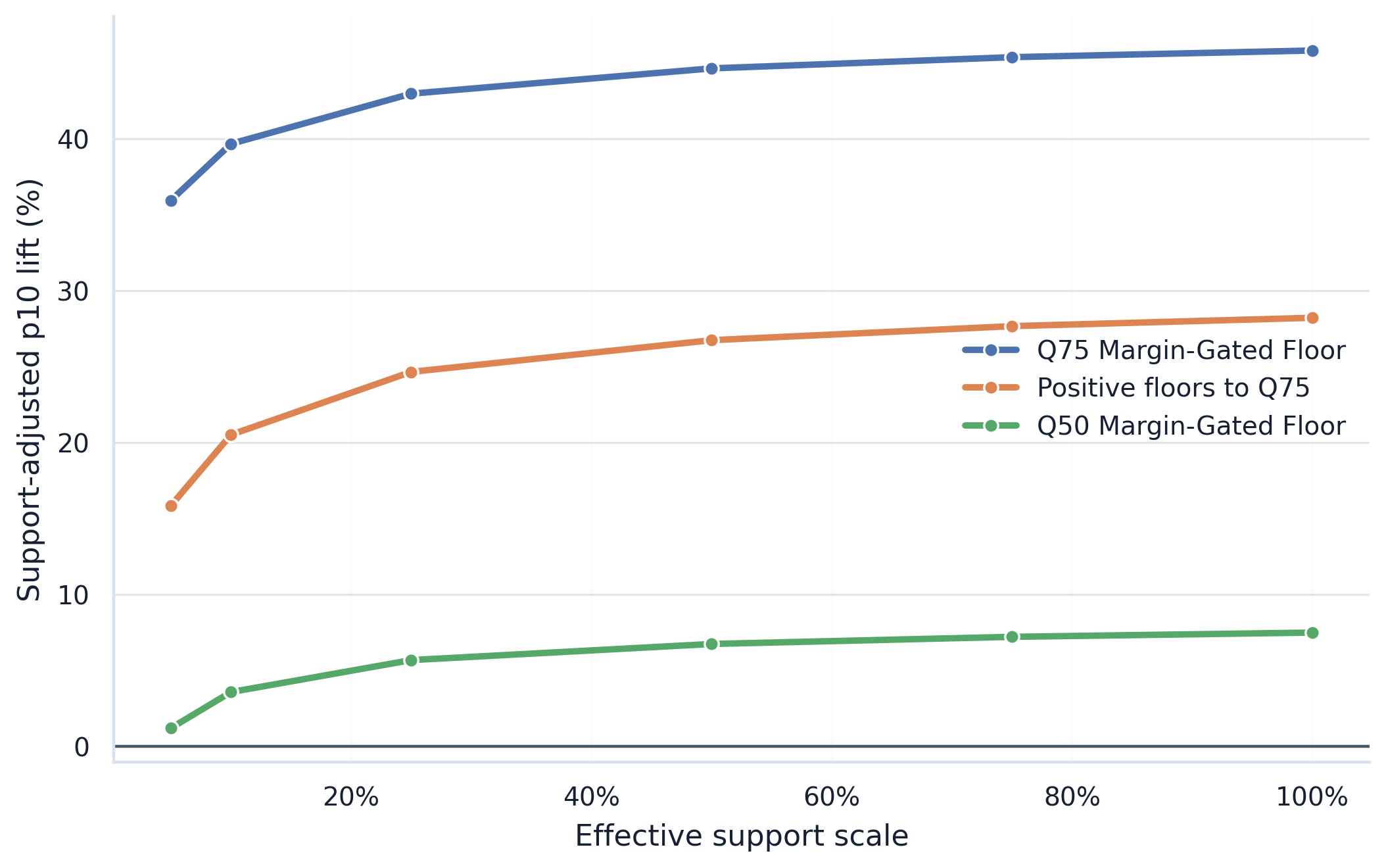}
\caption{\small Effective-support sensitivity. The curve starts from each policy's cross-fitted DR median and p10 lift, then widens the median-to-p10 gap by $1/\sqrt{s}$ as the effective support scale $s$ declines, as shown in Eq.~\eqref{eq:support_sensitivity}. The underlying artifact evaluates multiple shortlisted policies at six support scales. \prioritypolicy{} remains positive even under severe support contraction, but the declining lower-tail curve motivates shadow logging and controlled validation rather than immediate launch.}
\label{fig:support_sensitivity}
\end{figure}

Figure~\ref{fig:support_sensitivity} connects the OPE results to the support diagnostics. The priority policy's p10 DR lift is $45.8\%$ at full observed support. Under the support-contraction stress test, this lower-tail estimate declines but remains positive across the evaluated support scales. This strengthens the offline case that the recommendation is not driven only by a point estimate. At the same time, the slope of the curve reinforces the operational constraint. If production assignment probabilities are not logged or if the target policy moves into weakly supported regions, the model-assisted evidence becomes less reliable. Thus, support robustness improves the case for validation, not the case for skipping validation.

\begin{figure}[!t]
\centering
\includegraphics[width=0.78\linewidth]{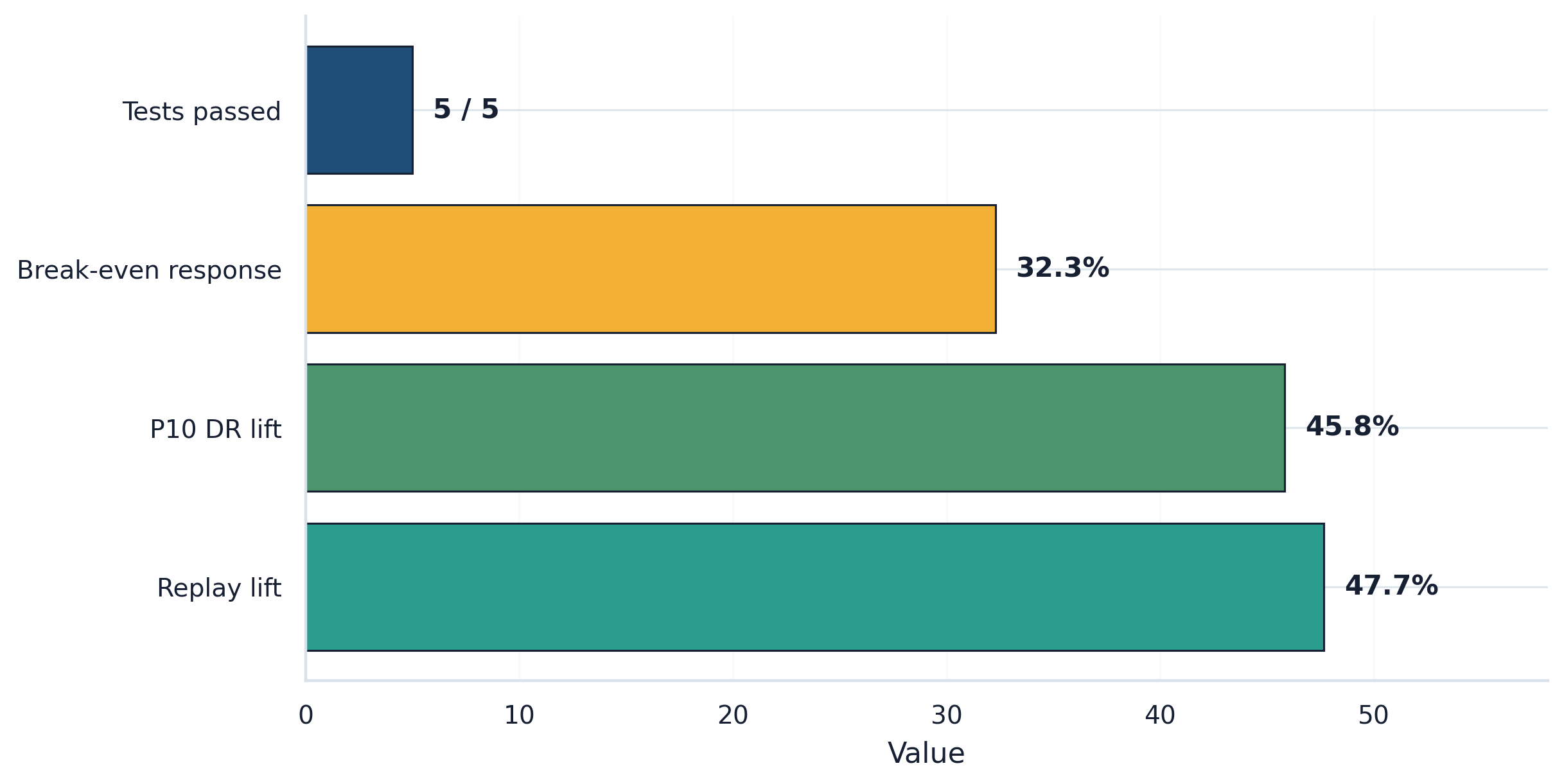}
\caption{\small Robustness summary for the priority policy. The summary combines replay lift, conservative p10 DR lift, break-even adverse response loss, and the number of robustness checks passed. \prioritypolicy{} has a 47.7\% replay lift, a 45.8\% conservative p10 DR lift, a 32.3\% break-even adverse response-loss share, and passes 5 out of 5 offline robustness checks. The decision remains validation-first because replay does not capture strategic marketplace response and OPE still depends on support.}
\label{fig:robustness_summary}
\end{figure}

Figure~\ref{fig:robustness_summary} summarizes the decision implication. The robustness experiments do not replace online evidence; they organize why online evidence is worth collecting. The ``5/5'' entry should be interpreted as a decision-support checklist rather than a set of formal hypothesis tests. The five passed checks are: (i) positive replay lift, (ii) positive conservative p10 DR lift, (iii) a substantial break-even adverse response-loss threshold, (iv) a positive lower-tail signal under the evaluated support-contraction scenarios, and a (v) validation-first recommendation that blocks direct launch. Thus, the priority policy passes all offline robustness checks used in this study, but the correct action is still shadow logging followed by a controlled switchback or randomized validation design. This distinction is central to the DSS framing.

The evidence up to this point supports validation, but not direct launch. The replay frontier in Fig.~\ref{fig:frontier} shows that \prioritypolicy{} delivers the strongest mechanical yield improvement, while the guardrail matrix in Fig.~\ref{fig:guardrail_matrix} shows that this gain does not come from obvious replay failures in fill, clicks, conversions, value proxy, or daily stability. The OPE diagnostics in Fig.~\ref{fig:opecomparison} and the conservative lower-tail ranking in Fig.~\ref{fig:conservative} further show that the policy remains favorable under model-assisted evaluation, and the robustness checks in Figs.~\ref{fig:response_sensitivity}--\ref{fig:robustness_summary} show that the recommendation is not fragile to the evaluated response-loss and support-contraction scenarios. However, all of these results are still offline as replay holds bids fixed, OPE depends on support and nuisance models, and robustness curves are stress tests rather than observed marketplace response. \textit{The appropriate conclusion is therefore not immediate launch, but an out-of-time validation step before any online experiment. This can also be alternatively formalized using (\ref{dpi2}) in terms of $Q_{-I}(\pi)$ being 1 but $I(\pi)$ being 0, thereby resulting in $d(\pi)$ being 0. And therefore concluding with a `validate online' recommendation}.

% For reproducibility, this subsection is generated from three deterministic artifacts. The response experiment uses `reserve_policy_effects.csv` and writes `equilibrium_sensitivity.csv` with 186 rows, corresponding to six policies and 31 response-loss values. The support experiment uses `advanced_policy_conservative_ranking.csv` and writes `support_collapse_curve.csv` with 36 rows, corresponding to six policies and six support scales. The summary figure uses `theory_guided_sensitivity_verdict.csv`. No model is refit in this notebook; all calculations are deterministic transformations of replay and cross-fitted OPE outputs produced in the preceding notebooks.

\subsection{Out-of-time season-three validation}

\label{subsec:season3_validation}

The preceding analyses identify \prioritypolicy{} as the strongest validation candidate using Season 2 replay, guardrails, OPE, heterogeneity, and robustness diagnostics. To test whether this recommendation transfers beyond the development window, we next apply the same frozen policy catalog to Season 3. This is an out-of-time validation exercise. In essence, the positive-floor quantiles and policy definitions are still those selected from Season 2, and no policy is re-tuned using Season 3 outcomes.

For each season $s \in \{2,3\}$ and policy $\pi$, define replay lift relative to the logged-floor status quo as
\[
\widehat{\Delta}_{\pi}^{(s)}
=
\frac{\widehat{Y}_{\pi}^{(s)}}{\widehat{Y}_{0}^{(s)}} - 1,
\]
where $\widehat{Y}_{\pi}^{(s)}$ is replay yield per opportunity under policy $\pi$ in season $s$, and $\widehat{Y}_{0}^{(s)}$ is the corresponding logged-floor baseline. The transfer gap is then
\[
G_{\pi}
=
\widehat{\Delta}_{\pi}^{(3)}
-
\widehat{\Delta}_{\pi}^{(2)}.
\]

\begin{figure}[!t]
\centering
\includegraphics[width=0.95\linewidth]{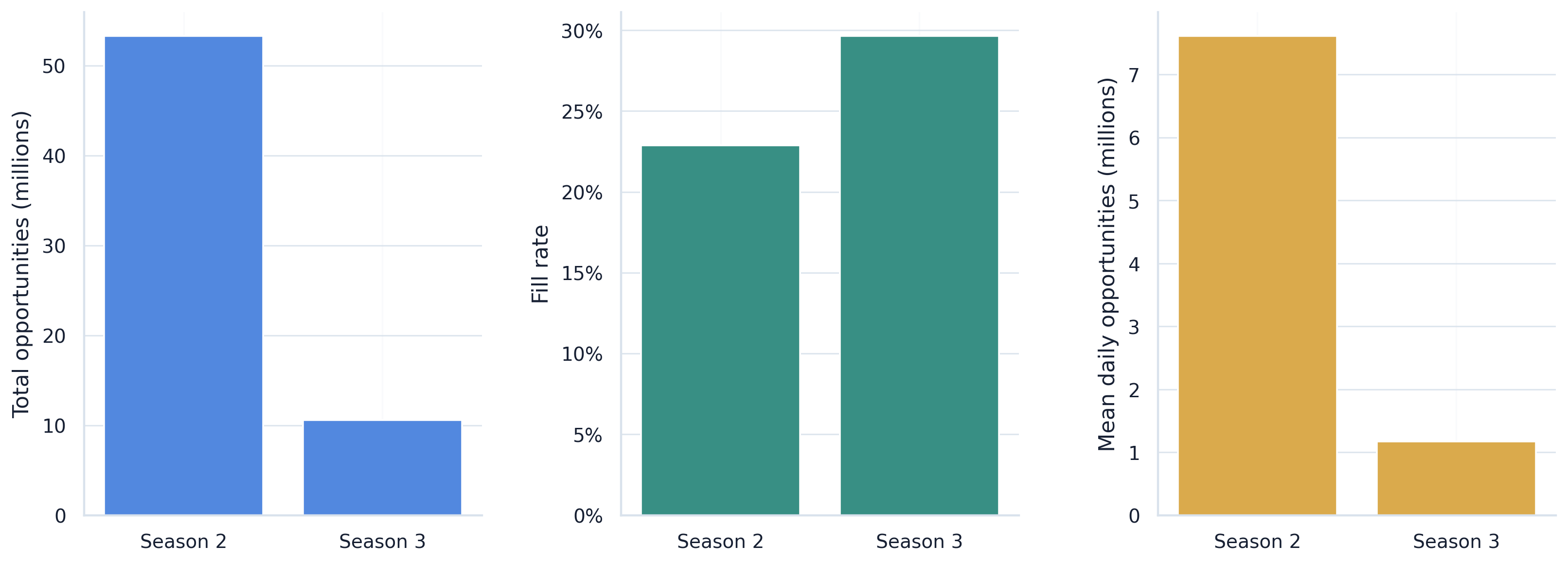}
\caption{\small Season 2 and Season 3 panel comparison. The validation window is smaller than the Season 2 development panel, with 10.57 million bid opportunities over 9 (Season 3) days compared with 53.29 million opportunities over 7 (Season 2) days. Season 3 has a higher fill rate, 29.6\% versus 22.9\%, and different daily scale.}
\label{fig:season3_distribution}
\end{figure}

Figure~\ref{fig:season3_distribution} establishes the data context for the holdout exercise. Season 3 is materially smaller in total opportunity volume and mean daily opportunity count, but it has a higher fill rate and higher event density. This matters because a policy that only works under the Season 2 traffic mix may not transfer. The holdout validation therefore asks a useful operational question that does the Season 2-selected policy remain attractive when the same replay contract is applied to a later, distributionally different panel?

\begin{figure}[!t]
\centering
\includegraphics[width=0.88\linewidth]{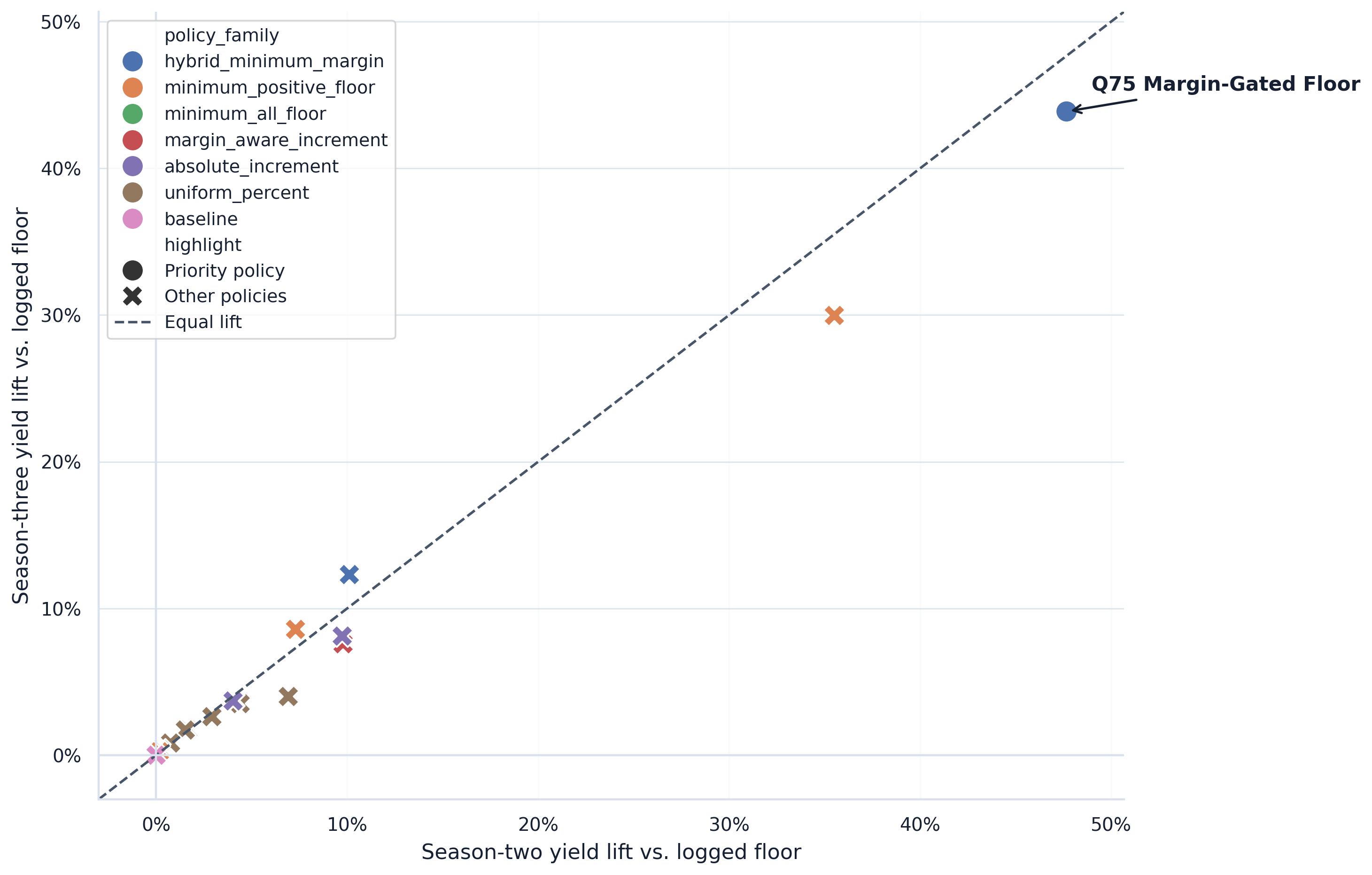}
\caption{\small Out-of-time transfer of replay lift from Season 2 to Season 3. Each point is a reserve/floor policy. The horizontal axis shows Season 2 replay lift and the vertical axis shows Season 3 replay lift, both relative to the logged-floor baseline within the same season. The dashed diagonal marks equal lift across seasons.}
\label{fig:season3_transfer}
\end{figure}

Figure~\ref{fig:season3_transfer} and Table~\ref{tab:season3transfer} report the main transfer result. \prioritypolicy{} remains the top-ranked policy in Season 3, with 43.9\% holdout replay lift compared with 47.7\% in Season 2. The rank is unchanged as it is rank 1 in both seasons. The table also shows that this transfer result does not come from sacrificing the retained-outcome proxies: Season 3 impression retention and value-proxy retention are both 100.00\% for the priority policy. The second-ranked policy also transfers directionally, but with a smaller Season 3 lift of 30.0\%. Thus, the holdout analysis approximately supports the same policy ordering that motivated the validation recommendation in the Season 2 analysis.

Table~\ref{tab:season3transfer} is useful because it makes the transfer result auditable across the full policy set, not only for the highlighted policy. Several lower-ranked policies shift modestly between seasons, but the leading candidates remain stable. In particular, the priority policy has a transfer gap of about $-3.8$ percentage points, which is small relative to its 47.7\% Season 2 lift and leaves a large 43.9\% Season 3 lift. This strengthens the interpretation that the recommendation is not an artifact of one development panel.

\begin{table}[!t]
\centering
\begin{threeparttable}
\caption{Out-of-time transfer from season two to season three.}
\label{tab:season3transfer}
\papertablesize
\setlength{\tabcolsep}{2.0pt}
\renewcommand{\arraystretch}{1.06}
\begin{tabularx}{\linewidth}{
  >{\raggedright\arraybackslash}p{0.07\linewidth}
  >{\raggedright\arraybackslash}p{0.27\linewidth}
  >{\raggedleft\arraybackslash}p{0.07\linewidth}
  >{\raggedleft\arraybackslash}p{0.07\linewidth}
  >{\raggedleft\arraybackslash}p{0.10\linewidth}
  >{\raggedleft\arraybackslash}p{0.10\linewidth}
  >{\raggedleft\arraybackslash}p{0.12\linewidth}
  >{\raggedleft\arraybackslash}p{0.12\linewidth}}
\toprule
Policy & Reader-facing policy & Season 2 rank & Season 3 rank & Season 2 lift & Season 3 lift & Season 3 impression retention & Season 3 value retention \\
\midrule
P18 & \prioritypolicy{} & 1 & 1 & 47.7\% & 43.9\% & 100.00\% & 100.00\% \\
P11 & Positive Floors To Q75 & 2 & 2 & 35.5\% & 30.0\% & 100.00\% & 100.00\% \\
P17 & Q50 Margin-Gated Floor & 3 & 3 & 10.1\% & 12.3\% & 100.00\% & 100.00\% \\
P13 & All Low Floors To Q50 & 3 & 3 & 10.1\% & 12.3\% & 100.00\% & 100.00\% \\
P10 & Positive Floors To Q50 & 6 & 4 & 7.3\% & 8.6\% & 100.00\% & 100.00\% \\
P8 & Add 20 To All Floors & 5 & 5 & 9.7\% & 8.1\% & 99.95\% & 99.86\% \\
P16 & Gap 100 Add 20 & 4 & 6 & 9.8\% & 7.6\% & 100.00\% & 100.00\% \\
P5 & Uniform +30\% & 7 & 7 & 6.9\% & 4.0\% & 99.40\% & 98.82\% \\
P15 & Gap 50 Add 10 & 9 & 8 & 4.1\% & 3.7\% & 100.00\% & 100.00\% \\
P7 & Add 10 To All Floors & 10 & 9 & 4.0\% & 3.7\% & 99.98\% & 99.96\% \\
P4 & Uniform +20\% & 8 & 10 & 4.4\% & 3.5\% & 99.87\% & 99.53\% \\
P3 & Uniform +15\% & 11 & 11 & 2.9\% & 2.6\% & 99.91\% & 99.76\% \\
P14 & Gap 25 Add 5 & 12 & 12 & 1.7\% & 1.7\% & 100.00\% & 100.00\% \\
P6 & Add 5 To All Floors & 13 & 13 & 1.7\% & 1.7\% & 99.99\% & 99.97\% \\
P2 & Uniform +10\% & 14 & 14 & 1.5\% & 1.7\% & 99.94\% & 99.85\% \\
P1 & Uniform +5\% & 15 & 15 & 0.8\% & 0.8\% & 99.97\% & 99.91\% \\
P12 & All Low Floors To Q25 & 16 & 16 & 0.2\% & 0.4\% & 100.00\% & 100.00\% \\
P9 & Positive Floors To Q25 & 17 & 17 & 0.2\% & 0.2\% & 100.00\% & 100.00\% \\
P0 & Logged Status Quo & 18 & 18 & 0.0\% & 0.0\% & 100.00\% & 100.00\% \\
\bottomrule
\end{tabularx}
\begin{tablenotes}
\papertablesize
\item Season-two and season-three lifts are replay yield lifts relative to the logged floor status quo within each season. Season-three retention columns report the share of baseline filled impressions and value proxy retained by the candidate policy in the season-three replay.
\end{tablenotes}
\end{threeparttable}
\end{table}

\begin{figure}[!t]
\centering
\includegraphics[width=0.98\linewidth]{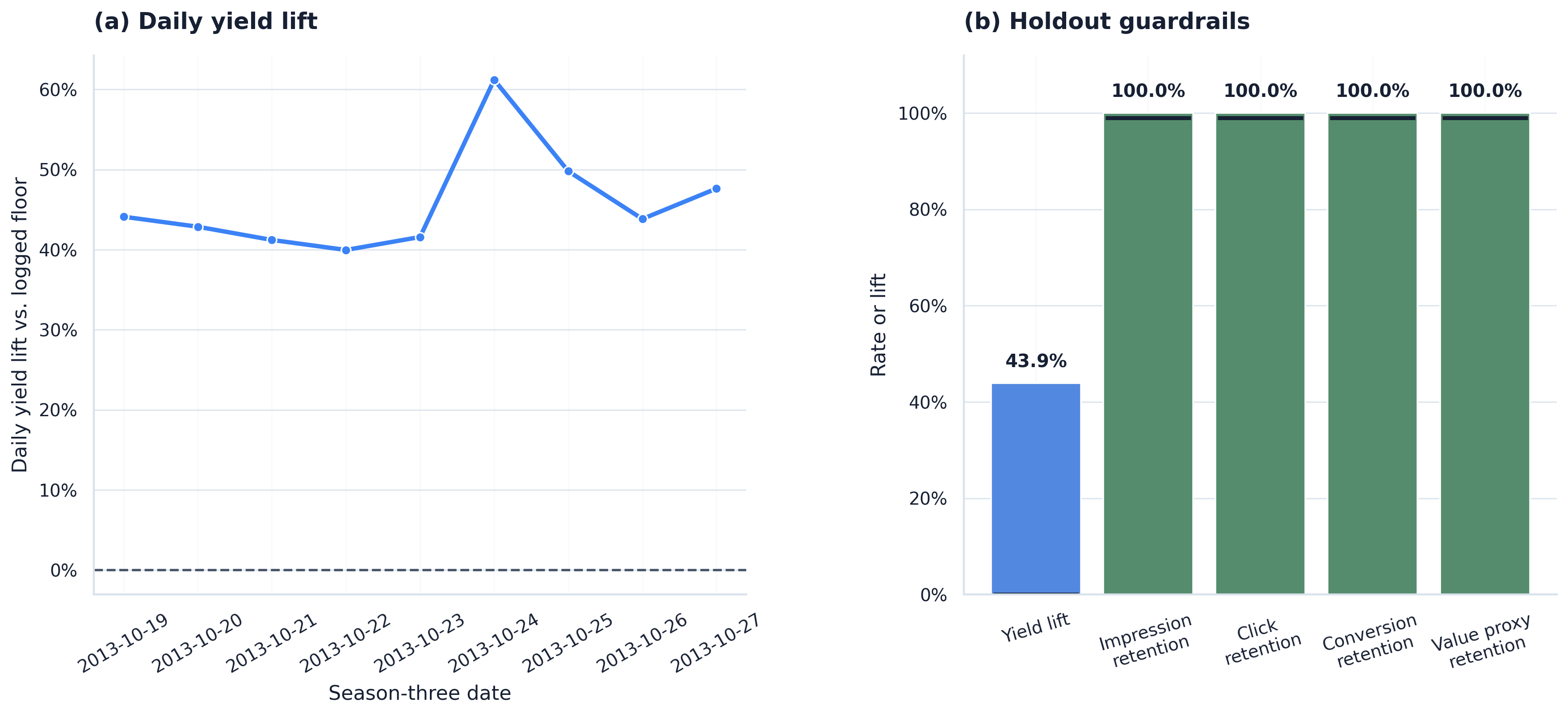}
\caption{\small Season 3 validation for the priority policy. Panel (a) reports daily Season 3 replay lift for \prioritypolicy{} relative to the logged-floor baseline. The lift is positive on all 9 holdout days, ranging from 40.0\% to 61.2\%. Panel (b) summarizes holdout guardrail quantities: Season 3 yield lift, impression retention, click retention, conversion retention, and value-proxy retention.}
\label{fig:season3_priority_validation}
\end{figure}

Figure~\ref{fig:season3_priority_validation} checks whether the aggregate Season 3 result is driven by a single unusually favorable day, or it has an acceptable support. It should be noted that the daily replay lift is positive on every Season 3 validation day, with a minimum daily lift of 40.0\%, a median daily lift of 43.8\%, and a maximum daily lift of 61.2\%. The paired guardrail panel is consistent with Table~\ref{tab:season3transfer} stating that the priority policy preserves the baseline impressions and value proxy in the Season 3 replay. This supports not only average lift, but also day-level stability and retained-outcome safety under the replay contract.

The interpretation remains deliberately conservative. Season 3 validation is stronger than within-season replay because it uses an out-of-time panel and a frozen policy catalog. However, it is still not a randomized online experiment. The replay contract continues to hold bids and participation fixed, and therefore does not identify bidder response, budget pacing response, or marketplace equilibrium effects. The correct conclusion is that Season 3 validation substantially strengthens the case for online validation of \prioritypolicy{}, not that the policy should be launched directly without a controlled experiment.

\subsection{Validation design and launch-readiness decision}
\label{subsec:validation_design_launch_readiness}

\begin{figure}[!t]
\centering
\includegraphics[width=0.98\linewidth]{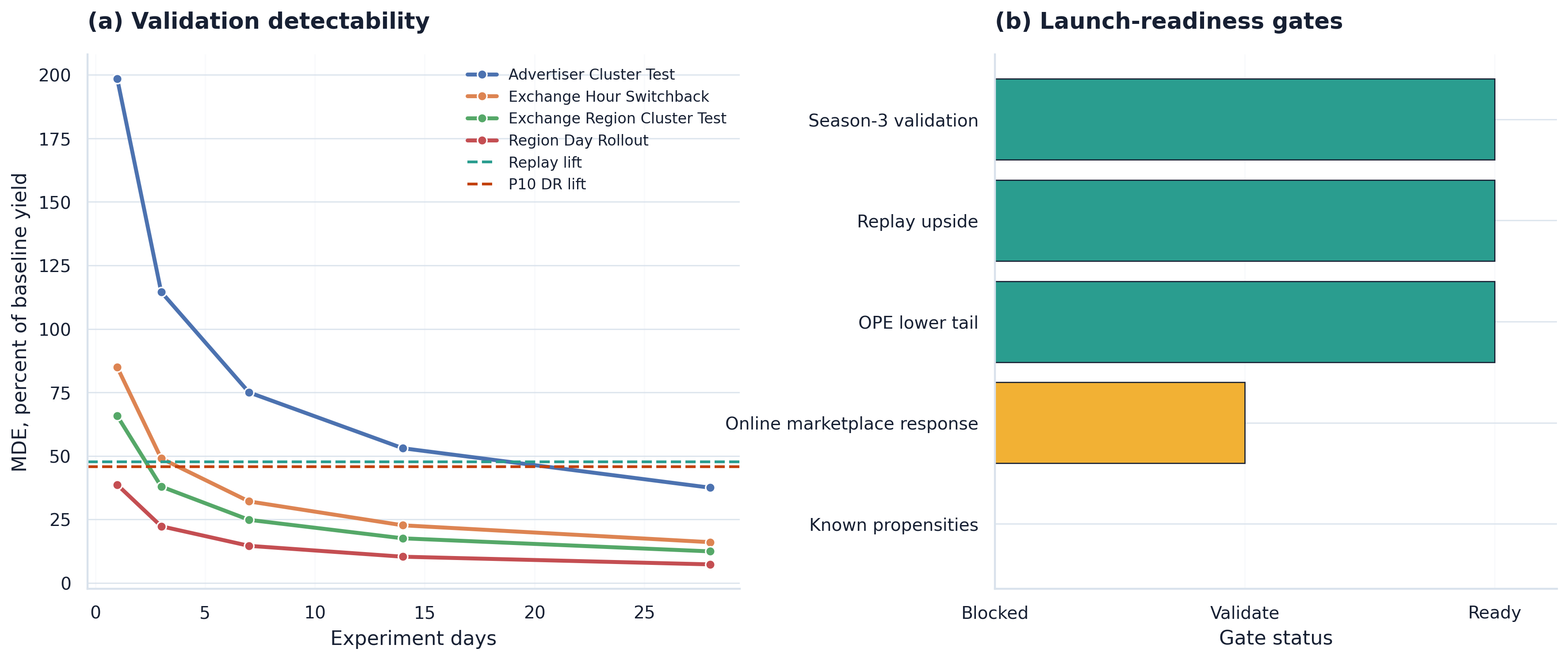}
\caption{\small Validation design and launch-readiness gates. Panel (a) reports data-calibrated MDE curves for candidate online validation designs. The dashed horizontal lines show the priority policy's Season 2 replay lift and conservative p10 DR lift. Lower curves indicate designs that can detect smaller effects over the same duration. Panel (b) summarizes launch-readiness gates. Replay upside, OPE lower-tail evidence, and Season 3 validation are ready; online marketplace response is validation-ready; known propensities remain blocked until real assignment probabilities are logged.}
\label{fig:validation_design_launch_readiness}
\end{figure}

The Season 3 holdout analysis strengthens the case for \prioritypolicy{}, but it still does not justify direct launch. The remaining uncertainty is whether the policy remains favorable after marketplace participants observe changed floors and respond through bids, budgets, pacing, or demand substitution. The final DSS step therefore converts the accumulated offline evidence into a validation-design and launch-readiness recommendation. For this purpose, we use the Minimum Detectable Effect (MDE) to quantify how large a live treatment effect must be for a proposed validation design to detect it reliably within a given experiment duration. In other words , we use the following analysis as a planning diagnostic, not as a guarantee of live effects. Let $\widehat{Y}_0$ denote logged-floor baseline yield per opportunity:
\[
\widehat{Y}_0
=
\frac{\sum_{i=1}^{n}Y_i^{(0)}}{n}.
\]
From the Season 2 logged-status-quo replay artifact,
$
n=53{,}289{,}330$, $
\sum_i Y_i^{(0)}=953{,}877{,}620,
$
so
$
\widehat{Y}_0 = 17.89997.
$
The priority policy's offline reference effects are the Season 2 replay lift and the conservative p10 cross-fitted DR lift:
$
\widehat{\Delta}_{\mathrm{replay}}=47.7\%,
$
$\widehat{\Delta}_{\mathrm{p10DR}}=45.8\%.
$
These appear as horizontal reference lines in Fig.~\ref{fig:validation_design_launch_readiness}(a). For validation design $d$ and duration $T$, the detectable effect is
\[
\mathrm{MDE}_d(T)=\frac{\widehat{c}_d}{\sqrt{T}},
\qquad
\mathrm{MDE}^{\mathrm{abs}}_d(T)=\widehat{Y}_0\mathrm{MDE}_d(T).
\]
Here $\widehat{c}_d$ is not assumed exogenously. It is estimated from Season 2 assignment-unit variation. For each design, we aggregate logged baseline yield and fill to the design's one-day assignment unit, compute the cluster-level standard deviation, and use the two-sample normal approximation
\[
\mathrm{MDE}^{\mathrm{abs}}_d(1)
=
(z_{0.975}+z_{0.80})
\frac{2\widehat{\sigma}_d}{\sqrt{G_d}},
\qquad
\widehat{c}_d
=
\frac{\mathrm{MDE}^{\mathrm{abs}}_d(1)}{\widehat{Y}_0}.
\]

where $G_d$ is the number of assignment units and $\widehat{\sigma}_{d}$ is the standard deviation of assignment-unit yield per opportunity. The assignment units are advertiser, exchange-hour, exchange-region, and region-day cells for the four designs.

Thus, for a 14-day exchange-hour switchback,
\[
\mathrm{MDE}_{\mathrm{switchback}}(14)
=
\frac{\widehat{c}_{\mathrm{switchback}}}{\sqrt{14}},
\qquad
\mathrm{MDE}^{\mathrm{abs}}_{\mathrm{switchback}}(14)
=
17.89997 \times \mathrm{MDE}_{\mathrm{switchback}}(14).
\]

The data-estimated $\widehat{c}_d$ values are used to draw the MDE curves as shown in Fig. \ref{fig:validation_design_launch_readiness}. Here panel (a) shows whether each design becomes sensitive enough to detect effects in the offline range within a practical window. The exchange-hour switchback is preferred because it balances detectability with marketplace validity as it avoids pure row-level randomization, which may be fragile under bidder budgets and pacing interference, while still providing repeated treated and control periods.

Panel (b) explains why validation is still required. Three evidence gates are ready: replay upside, OPE lower-tail evidence, and Season 3 out-of-time validation. But, the online marketplace-response gate is marked validation-ready because it can only be measured after changing serving behavior under a controlled design. The known-propensity gate is blocked because the public logs do not contain real randomized policy assignment probabilities. This is an important distinction as the DSS is not saying that the policy is weak. Rather, it is saying that the policy is strong enough to validate, while the remaining unresolved quantities are exactly the quantities offline replay cannot identify.

\begin{figure}[!t]
\centering
\includegraphics[width=0.98\linewidth]{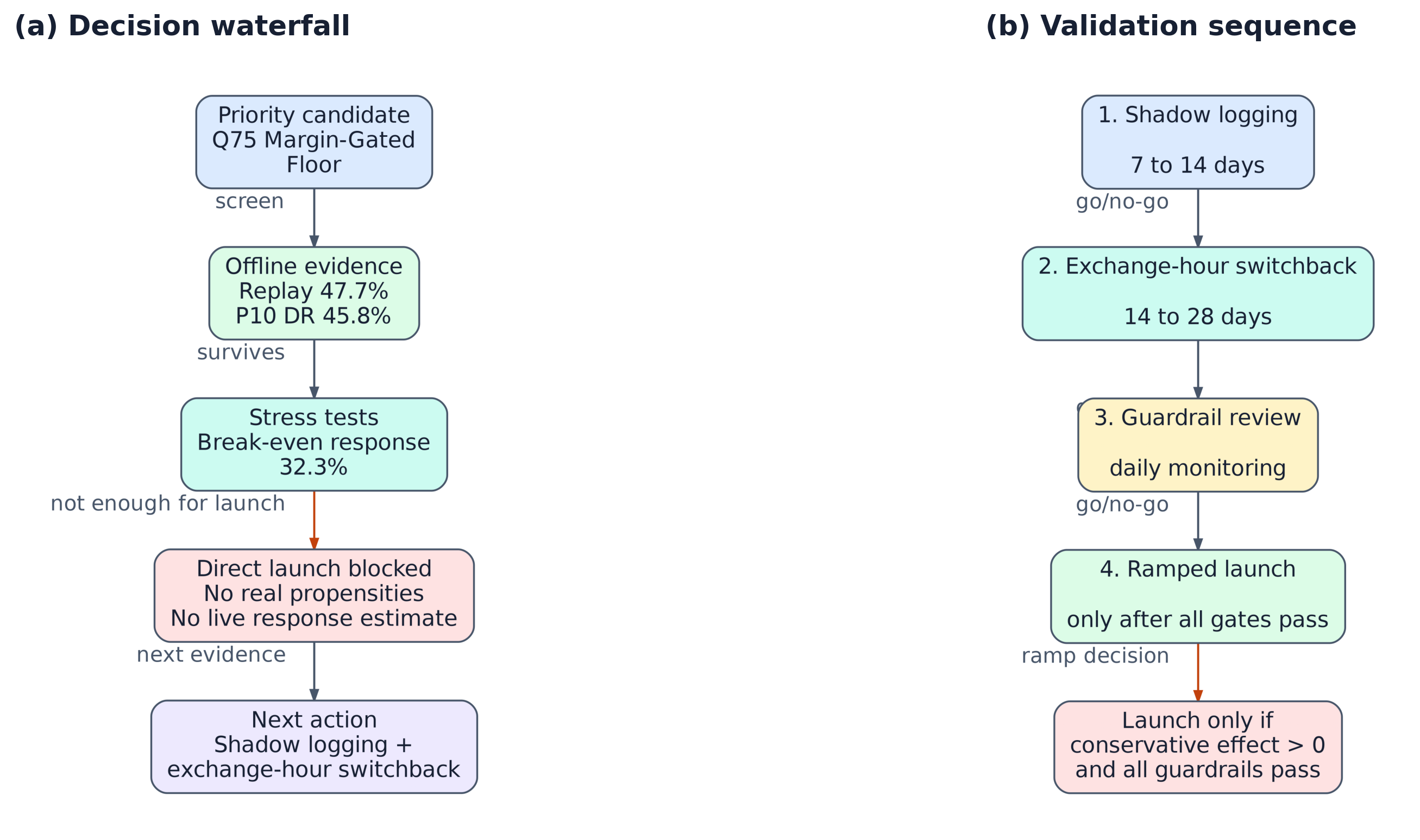}
\caption{\small Final DSS recommendation and validation sequence. Panel (a) shows the decision waterfall where \prioritypolicy{} is the priority candidate, the offline evidence is favorable, response-stress tests are passed, direct launch is blocked by missing real propensities and unvalidated live marketplace response, and the recommended action is shadow logging followed by an exchange-hour switchback. Panel (b) gives the implementation sequence starting with shadow logging for 7--14 days, then exchange-hour switchback for 14--28 days, daily guardrail review, and ramped launch only after all gates pass.}
\label{fig:decision_validation_sequence}
\end{figure}

Figure~\ref{fig:decision_validation_sequence} turns the statistical evidence into an operational recommendation. Panel (a) is the decision waterfall. The policy first enters as the priority candidate because it leads the replay frontier, passes replay guardrails, has favorable OPE lower-tail evidence, survives robustness checks, and transfers to Season 3. The waterfall then makes the launch blocker explicit by mentioning the evidence is still offline or replay-based, and therefore does not resolve real assignment propensities or live marketplace response.

Based on the complete analysis and MDE calculation, panel (b) makes the recommendation reproducible as a staged validation plan. The first stage is shadow logging for 7--14 days. During this phase, the system should log the current floor, candidate floor, bid price, whether the candidate floor would have cleared, the eligible policy set, and the assignment probability that would be used in a later experiment. This stage measures support and instrumentation quality without changing marketplace behavior. The second stage is an exchange-hour switchback for 14--28 days, which alternates treatment status across exchange-hour cells to reduce contamination from shared bidder budgets and pacing systems. The third stage is daily guardrail review, using fill, yield, clicks, conversions, value proxy, and segment-level diagnostics. Only after these stages pass should the system consider a ramped launch.

The final conclusion is therefore deliberately conservative. \prioritypolicy{} is not rejected as the empirical evidence is strong enough to justify validation. It is also not directly launched however as the unresolved gates correspond to causal and marketplace-response quantities that cannot be settled by static logs. This is the central DSS contribution of the paper. The framework does not merely rank policies by offline lift; it converts replay, guardrails, OPE, robustness, and holdout validation into an auditable launch-readiness recommendation.

\subsection{Decision-rule ablation}

\label{subsec:decision_ablation}

The preceding analyses show that \prioritypolicy{} is the strongest offline candidate, but they do not by themselves establish that the policy should be launched directly. This subsection isolates the value of the decision-support layer by comparing the full DSS against simplified decision rules that use only part of the evidence. The ablation asks two questions. First, would simpler rules identify the same leading policy? Second, would they make the same operational recommendation? This distinction is important because a rule can be correct about which policy is promising while still being too aggressive about launch readiness.

Table~\ref{tab:decisionablation} reports the ablation. Each row corresponds to a decision rule with a restricted information set. The replay-only rule selects the policy with the largest season-two replay lift. The replay-plus-guardrails rule adds the basic replay guardrails. The OPE rules use either the cross-fitted doubly robust mean lift or the conservative lower-tail lift from the OPE analysis. The season-three replay-only rule uses the frozen-policy holdout replay result from Table~\ref{tab:season3transfer}. The full DSS is allowed to use the complete evidence set, which includes replay, guardrails, OPE, support diagnostics, season-three transfer, response sensitivity, and interference/propensity launch gates.

\begin{table}[!t]
\centering
\begin{threeparttable}
\caption{Decision-rule ablation: simplified rules select the same policy but overclaim launch readiness.}
\label{tab:decisionablation}
\papertablesize
\setlength{\tabcolsep}{2.0pt}
\renewcommand{\arraystretch}{1.12}
\begin{tabularx}{\linewidth}{
  >{\raggedright\arraybackslash}p{0.17\linewidth}
  >{\raggedright\arraybackslash}p{0.25\linewidth}
  >{\raggedright\arraybackslash}p{0.17\linewidth}
  >{\raggedright\arraybackslash}p{0.13\linewidth}
  >{\centering\arraybackslash}p{0.09\linewidth}
  >{\centering\arraybackslash}p{0.11\linewidth}}
\toprule
Decision rule & Evidence used & Selected policy & Rule action & Launch overclaim & Unresolved direct-launch gates \\
\midrule
Replay-only & Season-two replay lift & P18: \prioritypolicy{} & Direct launch & Yes & 6 \\
Replay + guardrails & Replay lift and basic guardrails & P18: \prioritypolicy{} & Direct launch & Yes & 5 \\
OPE mean-only & Cross-fitted DR mean lift & P18: \prioritypolicy{} & Direct launch & Yes & 6 \\
OPE lower-tail-only & Conservative DR p10 lift and support & P18: \prioritypolicy{} & Direct launch & Yes & 5 \\
Season-three replay-only & Frozen-policy holdout replay lift & P18: \prioritypolicy{} & Direct launch & Yes & 5 \\
Full DSS & Replay, OPE, support, guardrails, holdout transfer, response, and interference gates & P18: \prioritypolicy{} & Online validation & No & 0 \\
\bottomrule
\end{tabularx}
\begin{tablenotes}
\papertablesize
\item Open gates are counted over seven evidence classes: replay, guardrails, OPE, support diagnostics, season-three validation, response sensitivity, and interference/propensity readiness. All simplified rules select the same priority policy, but they convert a favorable offline score into direct launch. The full DSS preserves the unresolved launch blockers and recommends online validation instead.
\end{tablenotes}
\end{threeparttable}
\end{table}

The main result is that all simplified rules select the same priority policy, P18: \prioritypolicy{}. This is reassuring that the recommendation is not an artifact of one scoring convention. Replay, OPE, and out-of-time validation all point to the same leading candidate. However, the simplified rules differ sharply from the full DSS in the action they recommend. The simplified rules stop after selecting a favorable policy and therefore imply direct launch. The full DSS instead recommends online validation. This is not a contradiction, rather it is the central decision-support result. Offline evidence is strong enough to justify validation capacity, but not sufficient to remove launch blockers associated with real propensities, bidder response, and marketplace interference.

\begin{figure}[!t]
\centering
\includegraphics[width=0.7\linewidth]{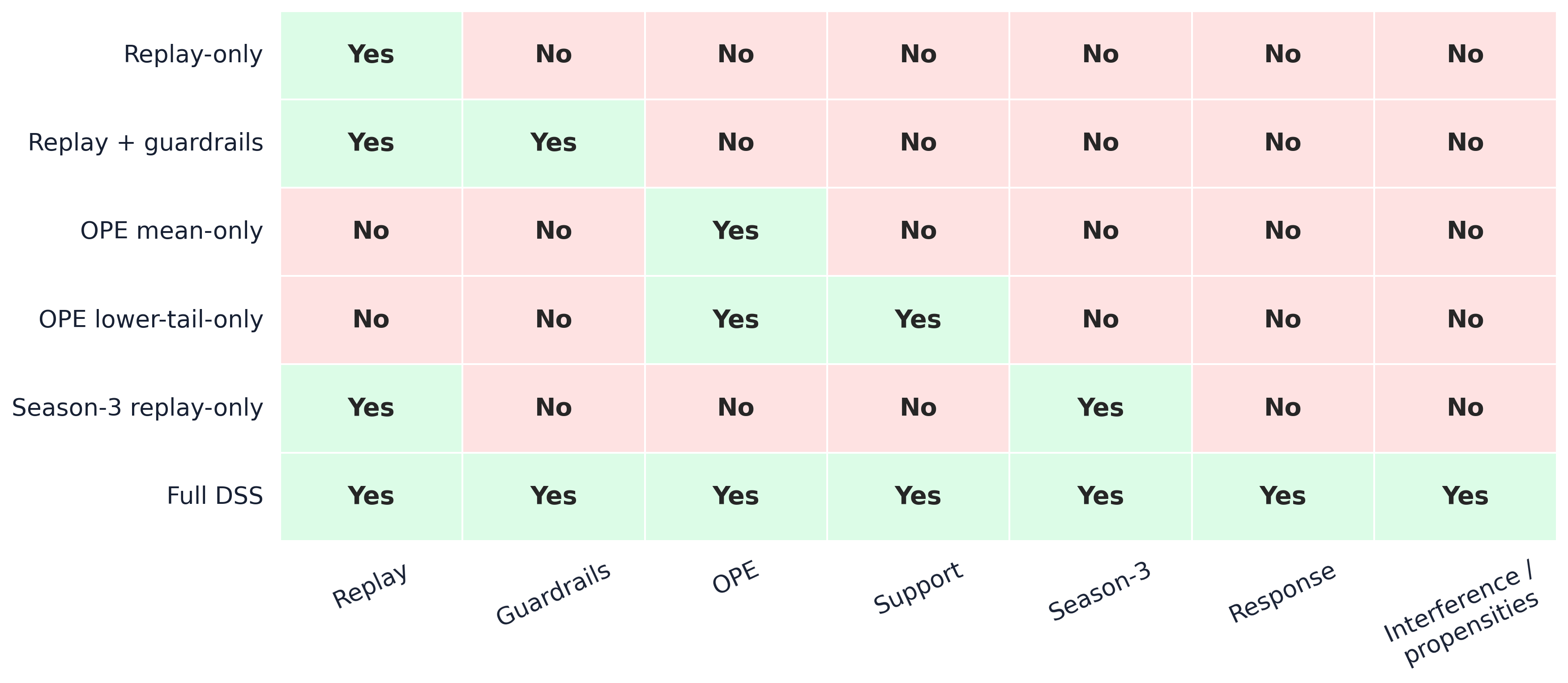}
\caption{\small Decision-rule evidence matrix. Each row is a candidate decision rule and each column is an evidence class used by the rule. Simplified rules use only a subset of the available evidence, whereas the full DSS combines replay, guardrails, OPE, support diagnostics, season-three validation, response sensitivity, and interference/propensity gates.}
\label{fig:ablation_gates}
\end{figure}

Figure~\ref{fig:ablation_gates} visualizes the information set behind each rule. The rows correspond to the decision rules in Table~\ref{tab:decisionablation}, and the columns correspond to evidence classes. The simplified rules each omit at least one major evidence class. Replay-only ignores guardrails, OPE, support, holdout transfer, response sensitivity, and interference/propensity readiness. OPE-only rules ignore the mechanical replay contract and external holdout transfer. Season-three replay-only uses out-of-time validation but does not use OPE uncertainty, support diagnostics, or response sensitivity. The full DSS is the only rule that keeps all evidence classes active at the decision point.

\begin{figure}[!t]
\centering
\includegraphics[width=0.6\linewidth]{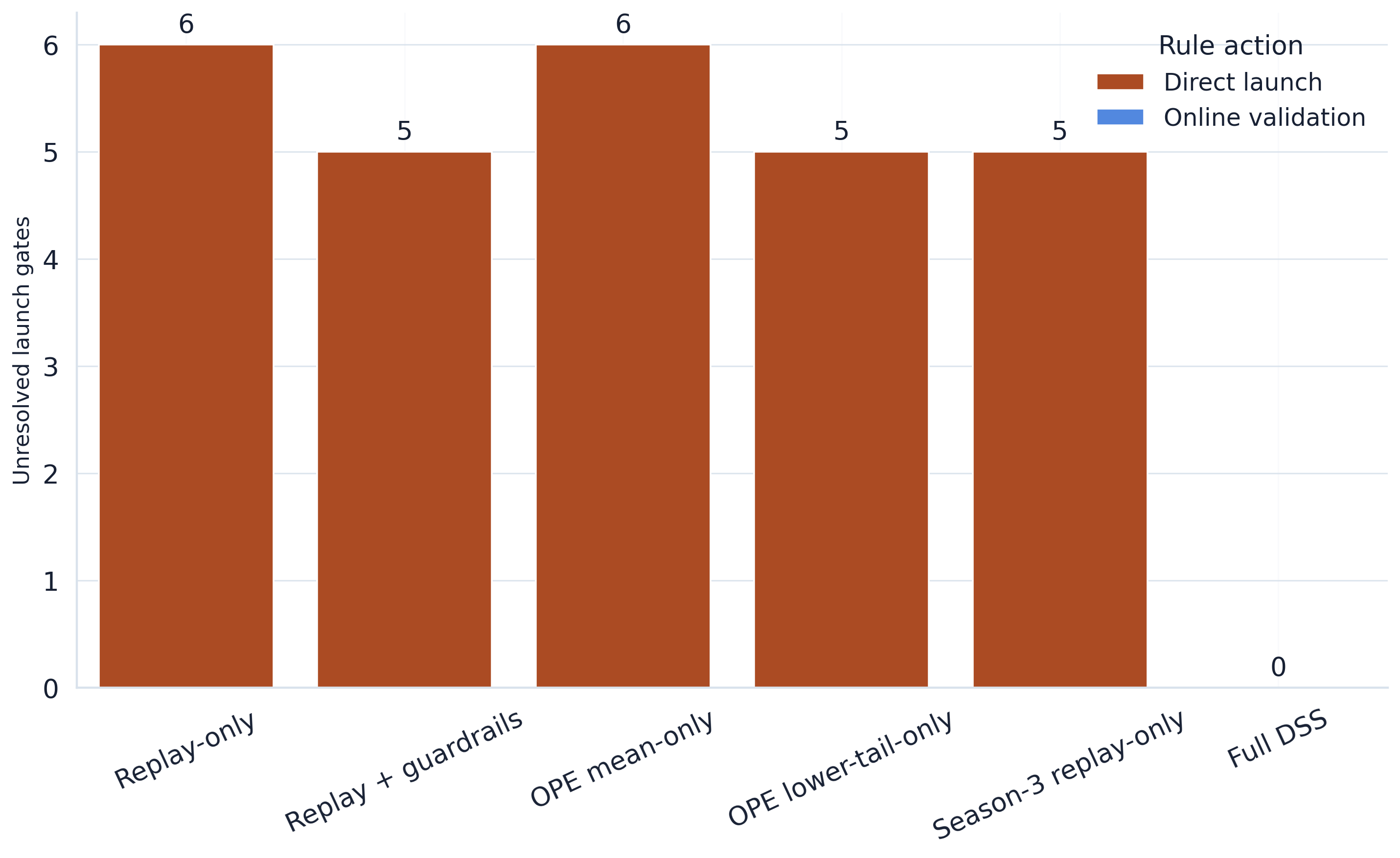}
\caption{\small Unresolved direct-launch gates under each decision rule. Simplified rules recommend direct launch despite omitting important evidence classes. The full DSS recommends online validation rather than launch, so it does not overclaim launch readiness.}
\label{fig:ablation_open_gates}
\end{figure}

Figure~\ref{fig:ablation_open_gates} converts the evidence matrix into the decision consequence. For simplified rules, the bar height is the number of unresolved direct-launch gates left open when that rule recommends launch. The replay-only and OPE mean-only rules each leave six of seven evidence classes unresolved. Replay-plus-guardrails and season-three replay-only leave five unresolved gates. OPE lower-tail-only also improves on OPE mean-only because it incorporates support diagnostics, but still leaves five direct-launch gates unresolved. In contrast, the full DSS has zero unresolved gates under its own recommended action because it does not recommend direct launch. It routes the policy to online validation, which is the appropriate action when offline evidence is favorable but live assignment and marketplace-response evidence are still missing.

\begin{figure}[!t]
\centering
\includegraphics[width=\linewidth]{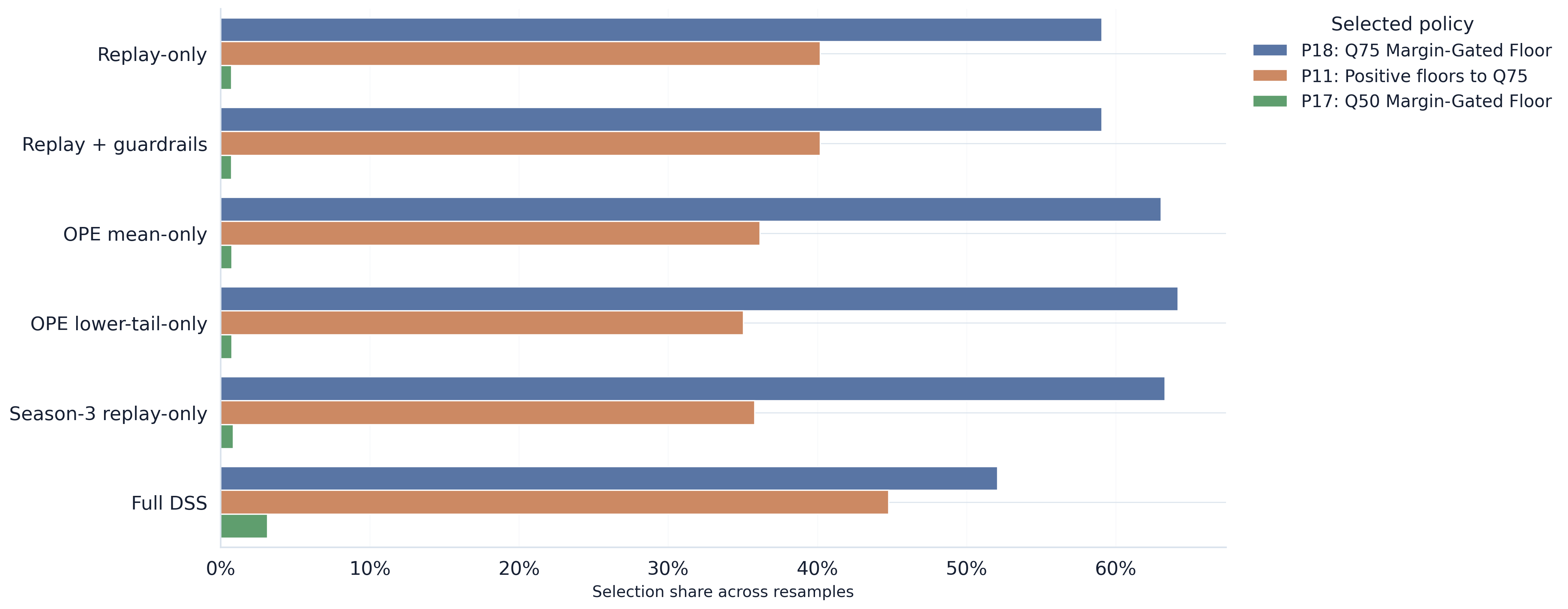}
\caption{\small Selection concentration across decision rules. Bars show the leading policies under each rule's scoring criterion. The priority policy remains the leading candidate across simplified and full DSS rules, but the full DSS changes the operational recommendation from direct launch to online validation.}
\label{fig:ablation_selection}
\end{figure}

Figure~\ref{fig:ablation_selection} shows the selection concentration for the leading policies under each rule. The figure is useful because it separates policy ranking from launch action. Across replay, OPE, season-three validation, and the full weighted DSS score, P18 remains the dominant candidate. P11, the positive-floors-to-Q75 policy, is usually the second strongest rule, while the Q50 margin-gated policy appears as a weaker third candidate. Thus, the ablation does not suggest that the full DSS is manufacturing a different winner by adding complexity. Instead, it shows that the full DSS changes the interpretation of the winner.

Taken together, Table~\ref{tab:decisionablation} and Figs.~\ref{fig:ablation_gates}--\ref{fig:ablation_selection} clarify the novelty of the framework. The contribution is not merely that replay, OPE, guardrails, and validation are individually useful. The contribution is the decision rule that keeps their distinct roles separate. Replay identifies mechanically plausible upside. Guardrails screen for basic retained-outcome safety. OPE adds model-assisted counterfactual evidence and uncertainty. Season-three replay tests transfer without retuning. Sensitivity analysis asks how much bidder response or support erosion the recommendation can tolerate. The final launch gate prevents these offline findings from being misread as live causal evidence. The ablation shows why this integration matters.

\section{Limitations}
\label{sec:limitations}

This study has several limitations. First, the analysis is based on logged auction data rather than a live randomized reserve-price experiment. Auction replay provides a mechanically transparent estimate of what would have happened under candidate floors when the logged bid is held fixed, but it does not observe how bidders would respond if the reserve policy were deployed. This is especially important in advertising marketplaces, where buyers may adapt bids, budgets, pacing, and participation after repeated exposure to a new floor policy.

Second, the OPE analysis uses a simulated policy logger to study estimator behavior under known propensities. This is useful for diagnosing support, estimator agreement, and lower-tail robustness, but it is not equivalent to observing true production randomization probabilities. The paper therefore treats OPE as validation evidence, not as a substitute for an online experiment with logged assignment probabilities.

Third, the season-three validation is an out-of-time replay validation, not an online causal validation. It strengthens the evidence that the priority policy is not overfit to one season-two window, but it still preserves the replay assumption that bids are fixed. As a result, season-three transfer supports external stability of the offline policy ranking, not direct launch readiness.

Fourth, the analysis uses value proxies derived from available click and conversion outcomes. These proxies are useful for guarding against obvious downstream degradation, but they are not a complete representation of advertiser welfare, user experience, budget pacing, auction competition, or long-run marketplace health. A production DSS would need richer platform-specific business metrics and monitoring constraints.

Finally, the policy class is intentionally interpretable and conservative. The candidate policies are reserve/floor rules based on logged floors, floor quantiles, and bid-floor margins. More flexible learned policies could potentially produce larger gains, but they would also require stronger support diagnostics, more careful monotonicity and safety constraints, and more online validation.

\section{Conclusion}
\label{sec:conclusion}

This paper develops a decision-support framework for evaluating reserve-price policies in an advertising marketplace. The empirical analysis starts with auction replay, uses guardrails to screen mechanically feasible candidates, adds model-assisted OPE diagnostics, checks heterogeneous downside risk, evaluates robustness to response and support assumptions, and then tests the frozen recommendation on an out-of-time season-three window. Across these analyses, the priority policy, \prioritypolicy{}, remains the leading candidate as it delivers the largest season-two replay lift, remains favorable under conservative OPE diagnostics, and transfers to season three without retuning.

The main conclusion, however, is not simply that this policy has high offline upside. The decision-rule ablation shows why an integrated DSS is needed. Simplified rules select the same policy, but they overclaim launch readiness by converting favorable offline scores into direct launch recommendations. The full DSS preserves the distinction between policy attractiveness and launch readiness. It recommends shadow logging followed by a controlled online validation design, rather than immediate deployment. This distinction is the paper's central contribution. In marketplace experimentation, the hard decision is often not whether an offline policy looks promising, but whether the evidence is strong enough to justify launch, validation, or rejection. By making replay, OPE, support, holdout transfer, guardrails, response sensitivity, and interference gates visible in one decision layer, the proposed framework turns offline policy evaluation into an auditable operational recommendation.

\section*{Declaration of Competing Interest}
The author(s) declare no competing interests.

\section*{Funding}
This research did not receive any specific grant from funding agencies in the public, commercial, or not-for-profit sectors.

\section*{Data and Code Availability}
The empirical analysis uses the public iPinYou real-time bidding (RTB) benchmark logs \citep{zhang2014ipinyou}. 
The raw logs are not redistributed in the manuscript repository; users should obtain the data from the original distributor and comply with its terms. All code, policy definitions, experiment configurations, and reproducibility artifacts are available at:

\begin{center}
\url{https://github.com/p-shekhar/marketplace-policies-code.git}
\end{center}

The repository contains code required to reproduce the reported analyses once the raw data are placed in the documented directory structure. If required by the journal, a versioned archival snapshot (e.g., Zenodo DOI) can be provided.

\bibliographystyle{unsrtnat}
\bibliography{dss_references}

\end{document}